\documentclass[useAMS,usenatbib]{mn2e}

\usepackage{amsmath}
\usepackage{graphics}
\usepackage{float} 
\usepackage{array}
\usepackage{color} 
\definecolor{Darkgreen}{rgb}{0,0.4,0}
\definecolor{Darkred}{rgb}{0.8,0,0} 
\definecolor{Darkblue}{rgb}{0.0, 0.0, 0.55}
\usepackage[colorlinks=true,urlcolor=blue,linkcolor=Darkblue,filecolor=black,citecolor=Darkred]{hyperref}
\usepackage{graphicx}
\usepackage{natbib}
\usepackage[final]{pdfpages} 
\usepackage{hyperref}
\usepackage{pdflscape}
\usepackage{lscape}
\usepackage{rotating}
\usepackage[section]{placeins}
\usepackage{stfloats}
\usepackage{xcolor,colortbl}
\usepackage{stfloats}

\newcommand{\deriv}{\mathrm{d}}

\definecolor{Gray}{gray}{0.85}

\title[Clustering-based redshift]{CLUSTERING-BASED REDSHIFT ESTIMATION : APPLICATION TO VIPERS/CFHTLS }

\author[V.~Scottez et al.]{V.~Scottez$^{1}$\thanks{E-mail:
scottez@iap.fr}, Y.~Mellier$^{1,2}$, B.~R.~Granett$^{3}$, T.~Moutard$^{5}$, M.~Kilbinger$^{1,2}$,  M.~Scodeggio$^{4}$, \and B.~Garilli$^{4}$,  M.~Bolzonella$^{8}$, S.~de la Torre$^{5}$, L.~Guzzo$^{3}$,  U.~Abbas$^{6}$, C.~Adami$^{5}$, \and S.~Arnouts$^{5}$, D.~Bottini$^{4}$, E.~Branchini$^{9,21,22}$, A.~Cappi$^{8,16}$, O.~Cucciati$^{14,8}$, I.~Davidzon$^{5,8}$, \and A.~Fritz$^{4}$, P.~Franzetti$^{4}$, A.~Iovino$^{3}$, J.~Krywult$^{13}$, V.~Le Brun$^{5}$, O.~Le F\`evre$^{5}$, \and D.~Maccagni$^{4}$, K.~Ma{\l}ek$^{18}$, F.~Marulli$^{14,15,8}$, M.~Polletta$^{4}$, A.~Pollo$^{17,18}$, L.~A.~.M.~Tasca$^{5}$, \and R.~Tojeiro$^{10}$, D.~Vergani$^{19,8}$, A.~Zanichelli$^{20}$,   J.~Bel$^{7}$,  J.~Coupon$^{11}$, G.~De Lucia$^{12}$, \and O.~Ilbert$^{5}$, H.~J.~McCracken$^{1}$ and L.~Moscardini$^{14,15,8}$ \\
\\
~~~~~~~~~~~~~~~~~~~~~~~~~~~~~~~~~~~~~~~~~~\textit{\small(Affiliations can be found at the end of the paper)} }

\begin{document}

\def\aj{AJ}%
\def\actaa{Acta Astron.}%
\def\araa{ARA\&A}%
\def\apj{ApJ}%
\def\apjl{ApJ}%
\def\apjs{ApJS}%
\def\ao{Appl.~Opt.}%
\def\apss{Ap\&SS}%
\def\aap{A\&A}%
\def\aapr{A\&A~Rev.}%
\def\aaps{A\&AS}%
\def\azh{AZh}%
\def\baas{BAAS}%
\def\bac{Bull. astr. Inst. Czechosl.}%
\def\caa{Chinese Astron. Astrophys.}%
\def\cjaa{Chinese J. Astron. Astrophys.}%
\def\icarus{Icarus}%
\def\jcap{J. Cosmology Astropart. Phys.}%
\def\jrasc{JRASC}%
\def\mnras{MNRAS}%
\def\memras{MmRAS}%
\def\na{New A}%
\def\nar{New A Rev.}%
\def\pasa{PASA}%
\def\pra{Phys.~Rev.~A}%
\def\prb{Phys.~Rev.~B}%
\def\prc{Phys.~Rev.~C}%
\def\prd{Phys.~Rev.~D}%
\def\pre{Phys.~Rev.~E}%
\def\prl{Phys.~Rev.~Lett.}%
\def\pasp{PASP}%
\def\pasj{PASJ}%
\def\qjras{QJRAS}%
\def\rmxaa{Rev. Mexicana Astron. Astrofis.}%
\def\skytel{S\&T}%
\def\solphys{Sol.~Phys.}%
\def\sovast{Soviet~Ast.}%
\def\ssr{Space~Sci.~Rev.}%
\def\zap{ZAp}%
\def\nat{Nature}%
\def\iaucirc{IAU~Circ.}%
\def\aplett{Astrophys.~Lett.}%
\def\apspr{Astrophys.~Space~Phys.~Res.}%
\def\bain{Bull.~Astron.~Inst.~Netherlands}%
\def\fcp{Fund.~Cosmic~Phys.}%
\def\gca{Geochim.~Cosmochim.~Acta}%
\def\grl{Geophys.~Res.~Lett.}%
\def\jcp{J.~Chem.~Phys.}%
\def\jgr{J.~Geophys.~Res.}%
\def\jqsrt{J.~Quant.~Spec.~Radiat.~Transf.}%
\def\memsai{Mem.~Soc.~Astron.~Italiana}%
\def\nphysa{Nucl.~Phys.~A}%
\def\physrep{Phys.~Rep.}%
\def\physscr{Phys.~Scr}%
\def\planss{Planet.~Space~Sci.}%
\def\procspie{Proc.~SPIE}%
\let\astap=\aap
\let\apjlett=\apjl
\let\apjsupp=\apjs
\let\applopt=\ao
\uchyph=0

\date{Accepted June 2016 ; Received May 2016 ; in original form May 2016}

{\pagerange{\pageref{firstpage}--\pageref{lastpage}} \pubyear{2016}

\maketitle

\label{firstpage}

\begin{abstract}

  \noindent We explore the accuracy of the clustering-based redshift
  estimation proposed by \cite{menard_2013} when applied to VIPERS and
  CFHTLS real data. This method enables us to reconstruct redshift
  distributions from measurement of the angular clustering of objects
  using a set of secure spectroscopic redshifts. We use state-of-the
  art spectroscopic measurements with $i_{\text{AB}}<22.5$ from the
  VIMOS Public Extragalactic Redshift Survey (VIPERS) as
  reference population to infer the redshift distribution of
  galaxies from the Canada-France-Hawaii Telescope Legacy Survey
  (CFHTLS) T0007 release. VIPERS provides a nearly representative
  sample to a flux limit of $i_{\text{AB}}<22.5$ at a redshift of
  $>0.5$ which allows us to test the accuracy of the clustering-based
  redshift distributions. We show that this method enables us to
  reproduce the \textit{true} mean colour-redshift relation when both
  populations have the same magnitude limit. We also show that this
  technique allows the inference of redshift distributions for a
  population fainter than the reference and we give an
  estimate of the colour-redshift mapping in this case. This last point
  is of great interest for future large redshift surveys which require
  a complete faint spectroscopic sample.

\end{abstract}


\section{Introduction}
\label{sec:intro}

Large future redshift surveys like the ESA Euclid space mission
\citep{euclid_mission,amendola_2012} aim to probe dark energy with
unprecedented accuracy. Many of the cosmological measurements to be
performed with these surveys \-- e.g. tomographic weak lensing,
tomographic clustering \-- will require extremely well characterised
redshift distributions
\citep{albrecht_2006,huterer,ma_2006,thomas_2011}.

Since it is impractical to measure spectroscopic redshifts for
hundreds of millions of galaxies \-- especially extremely faint ones
\-- these experiments are largely dependent upon photometric
redshifts: i.e. estimates of the redshifts of objects based only on
flux information obtained through broadband filters. Photos-z also
require large spectroscopic samples both for the calibration of
empirical methods \citep{Connolly_1995} and the building of
representative template libraries for template-fitting techniques
\citep{coleman_1980}. However, current and future spectroscopic
surveys will be highly incomplete due to selection biases dependent on
redshift and galaxy properties \citep{cooper_2006}. Because of this,
along with the catastrophic photometric errors that can occur at a
significant ($\sim 1\%$) rate \citep{sun_2009,bernstein_2010}
photometric redshifts are not sufficiently precise. If future dark
energy experiments have to reach their goals it is necessary to
develop a method to infer, at least, the redshift distribution with
high precision.  

Current projections for cosmic shear
measurements estimate that the true mean redshift of objects in each
photo-z bin must be known to better than $\sim$ 0.002(1 +
$z$) \citep{zhan_and_knox_2006,zhan_2006,knox_2006} with stringent
requirements on the fraction of unconstrained catastrophic outliers
\citep{hearin_2006} while the width of the bin must be known to
$\sim$ 0.003(1 + $z$). \cite{newman_2013_spectro_needs} investigated
the spectroscopic needs for dark energy imaging experiments and
insisted on the extremely high ($\sim 99.9\% $) completeness required
for calibration techniques.  

The idea of measuring
redshift distributions using the apparent clustering of objects on the
sky is not new. It was first developed by \cite{seldner_peebles_1979};
\cite{phillipps_shanks_1987} and \cite{landy_szalay_1996}. This was
practically forgotten mainly due to the rise of photometric
redshifts. To face the challenges of future and ongoing dark energy
imaging experiments \cite{newman_2008}, \cite{mattew_newman_2010} and
\cite{matthews_newman_2012} reapplied this method on simulations
while \cite{mcquinn_white_2013} proposed an optimal estimator for such
a measurement.  In this paper, we explore the clustering-based
redshift estimation, i.e \textit{cluster-z}, via a local (i.e. within
few Mpc) approach introduced by \cite{menard_2013} (M13), validated
with simulations by \cite{schmidt} and compared to spectroscopic
redshift at limiting magnitude $r_{\text{model}}<19$ by
\cite{rahman_2015} (R15). Recently \cite{schmidt_2015} applied this
technique to continuous fields by inferring the redshift distribution
of the cosmic infrared background while \cite{rahman_2015_IR} and
\cite{rahman_2015_sdssphotz} explored this method in near infrared
using 2MASS Extended and Point Source Catalogs as well as the SDSS
Photometric Galaxies. This work aims to explore the strength of
cluster-z at fainter magnitude $i_{\text{AB}}<22.5$ using
real data similar to what will be available with Euclid in term of
filters and observational strategy and demonstrate our ability to
recover the redshift distribution of an unknown sample with
$22.5<i_{\text{AB}}<23$ when the reference sample used for
calibration has only $i_{\text{AB}}<22.5$.

This paper is organised as follows. In Section~\ref{sec:clusterz_formalism}, we review the clustering-based
redshift formalism while the data used in this work are described in
Section~\ref{sec:data_analysis}. Then in
Section~\ref{sec:tomographic} we show our ability to measure the
clustering redshift distribution using a tomographic photo-z
approach. We also show that this method allows the estimation of
redshift distribution when the sample of unknown redshift is
fainter than the reference one. Finally we free
cluster-z from the use of photos-z in
Section~\ref{sec:colors_sampling} by selecting subsamples in
colour-space and we explore in this case the reconstruction of the
colour-redshift mapping for faint objects. Conclusions are presented
in Section~\ref{sec:concl}.

\section{Clustering-based redshift: formalism}
\label{sec:clusterz_formalism}

The method used in this paper is based on the work of M13 and R15. We
refer the reader to those papers for more details. In this section, we
briefly review the formalism.

The key point is that correlated galaxies are at the same location in
redshift and on the sky. Sources at different redshift are
uncorrelated. This clustering information is encoded into the
two-point correlation function as an excess probability \-- compared
to a random distribution \-- to find two objects close together. This
is valid in 3D and, by projection, on the sky. Using a
reference sample of secure spectroscopic redshifts \-- and by
looking at the galaxy cluster scale \-- it is then possible to extract
the \textit{excess probability} of finding a population of galaxies at
a given redshift. Obviously the reference population and the
unknown one \-- for which angular positions are known but
redshifts are not \-- have to overlap on the sky.  \\~~

The mean surface density of unknown objects at a distance $\theta$ from a reference one which is at a redshift $z$, is:
\begin{equation}
\Sigma_{\text{ur}}(\theta,z)  = \Sigma_{\text{R}}  [1+ \omega_{\text{ur}}(\theta,z)] \ ,
\end{equation}
where $\Sigma_{\text{R}}$ is the random surface density of the unknown sample and $\omega_{\text{ur}}(\theta,z)$ is the two-point angular cross-correlation function between the two samples. Then, one can define the integrated cross-correlation function as:
\begin{equation}
\bar{\omega}_{\text{ur}}(z) = \int_{\theta_{\text{min}}}^{\theta_{\text{max}}} \deriv \theta \ W(\theta) \ \omega_{\text{ur}} (\theta,z) \ ,
\label{def_fonction_correlation_integre}
\end{equation}
where the range covered by $\theta$ varies with redshift and correspond to physical distances from few hundred kiloparsecs to several megaparsecs. Here we worked within a $[0.2;6]$ Mpc annulus. $W(\theta)$ is a weight function \-- $\propto \theta^{-0.8}$ \-- aimed at optimising the overall $S/N$ and whose integral is normalised to unity. This integrated cross-correlation function represents the excess probability, with respect to a Poisson distribution, to find an object of the unknown sample at an angular distance between $\theta_{\text{min}}$ and $\theta_{\text{max}}$ from a generic object of the reference sample at redshift $z$.

One can also write this quantity as a function of the redshift selection function of sample $\text{i} \in \left\lbrace \text{u},\text{r} \right\rbrace$ , $\deriv \text{N}_{\text{i}} /  \deriv z$, as well as the galaxy-dark matter biases, $\bar{b}_{\text{i}}(z)$, and the dark matter correlation function, $\bar{\omega}_{\text{m}}(z)$:
\begin{equation}
\bar{\omega}_{\text{ur}} = \int \deriv z' \  \frac{\deriv \text{N}_{\text{u}}}{\deriv z}(z') \ \frac{\deriv \text{N}_{\text{r}}}{\deriv z}(z') \ \bar{b}_{\text{u}}(z') \ \bar{b}_{\text{r}}(z') \ \bar{\omega}_{\text{m}}(z') \ .
\end{equation}
Applying the narrow sample approximation for the reference sample $\deriv \text{N}_{\text{r}} / \deriv z= \text{N}_{\text{r}}  \delta_{\text{D}} (z' - z)$ \-- with $\delta_{\text{D}}$ the Dirac delta function \-- we can then simply invert the previous integral and get:
\begin{equation}
\frac{\deriv \text{N}_{\text{u}}}{\deriv z}(z) \propto \bar{\omega}_{\text{ur}}(z) \times \frac{1}{\bar{b}_{\text{u}}(z)} \times \frac{1}{\bar{b}_{\text{r}}(z) \bar{\omega}_{\text{m}}(z)} \ ,
\label{full_n_u_z}
\end{equation}
where $\bar{\omega}_{\text{ur}}(z) $ can be directly measured in data, $\bar{b}_{\text{r}}(z)$ can be measured in the reference sample,  $\bar{\omega}_{\text{m}}(z)$ is given by the cosmology and $\bar{b}_{\text{u}}(z)$ is the only unknown quantity.

Considering a narrow redshift distribution for the unknown sample we can neglect the variation of its galaxy-dark matter bias with respect to the variation of the number of objects:
\begin{equation}
\frac{\deriv \log \  \deriv \text{N}_{\text{u}} / \deriv z }{\deriv z} \gg \frac{\deriv \log \ \bar{b}_{\text{u}}}{\deriv z} \ ,
\label{bias_variation}
\end{equation}
we get:
\begin{equation}
\frac{\deriv \text{N}_{\text{u}}}{\deriv z}(z) \propto  \bar{\omega}_{\text{ur}}(z) \left(  \frac{1}{\bar{b}_{\text{r}}(z) \ \bar{\omega}_{\text{m}}(z)} \right) \ .
\label{n_u_z_propto}
\end{equation}
As in Equation\eqref{bias_variation} we can neglect the redshift variation of $\sqrt{\bar{\omega}_{\text{m}}}$ with respect to $\deriv \text{N}_{\text{r}} / \deriv z$:
\begin{equation}
\frac{\deriv \log \ \deriv \text{N}_{\text{r}} / \deriv z }{\deriv z} \gg \frac{\deriv \log \ \sqrt{\bar{\omega}_{\text{m}}}}{\deriv z} \ .
\label{omegam_variation}
\end{equation}
Thus introducing the clustering amplitude of the reference sample, $\beta_{\text{r}} (z)$, we can write:
\begin{equation}
\beta_{\text{r}}(z)= \sqrt{\frac{\bar{\omega}_{\text{rr}}(z)}{\bar{\omega}_{\text{rr}}(z_{\text{0}})}} \propto       \frac{\bar{b}_{\text{r}}(z)}{\bar{b}_{\text{r}}(z_{\text{0}})} \ .
\label{clust_amp}
\end{equation}
Note that we can define $\beta_{\text{u}}(z)$ in the same way. As explained in (R15) one should note that this quantity is different from the linear “galaxy bias” which is usually defined only on large scales for which the galaxy and dark matter density fields are, on average, linearly related. This bias definition includes contributions from small scales where the galaxy and matter fields are non-linearly related.
We can then rewrite a model-independent version of Equation\eqref{n_u_z_propto} and we get:
\begin{equation}
\frac{\deriv \text{N}_{\text{u}}}{ \deriv z}(z) \propto  \bar{\omega}_{\text{ur}}(z) / \beta_{\text{r}}(z) \ .
\label{n_u_z_propto_beta}
\end{equation}
Finally, the redshift distribution is normalised to the number of objects in the unknown sample through:
\begin{equation}
\int \deriv z \ \deriv  \text{N}_{\text{u}} / \deriv z = \text{N}_{\text{u}} \ .
\label{normalisation}
\end{equation}
It is important to realise that to be able to write and use Equation\eqref{n_u_z_propto_beta} we have to select unknown samples with relatively small redshift distributions to have $\bar{b}_{\text{u}}(z)$ or $\beta_{\text{u}}(z)$ slowly varying with redshift. The ability of selecting subsamples with narrow redshift distributions is quite important to consider: $\deriv \beta_{\text{u}} / \deriv z = 0$. 

\noindent To directly measure the integrated cross-correlation function we can simply use the \cite{davis_peebles_1983} estimator:
\begin{equation}
\bar{\omega}_{\text{ur}}(z)= \frac{ \langle \Sigma_{\text{ur}}\rangle_z }{ \Sigma_{\text{R}} } -1 \ .
\label{estimateur_omega_bar}
\end{equation}
The error in the measurement is then estimated through Poisson statistic and is given by:
\begin{equation}
\sigma^2_{\bar{\omega}} = \left( \frac{\bar{\omega}+1}{\sqrt{\text{N}_{\text{ur}}}} \right)^2 +  \left( \frac{\bar{\omega}+1}{\sqrt{\text{N}_{\text{R}}}} \right)^2 \ ,
\end{equation}
where $\text{N}_{\text{ur}}$ is the neighbours number of unknown objects over $[\theta_{\text{min}};\theta_{\text{max}}]$ around reference galaxies and $\text{N}_{\text{R}}$ is the corresponding number of neighbours for a random distribution.

\section{Data analysis}
\label{sec:data_analysis}

\subsection{The Datasets}
\label{sec:datasets}

\subsubsection{VIPERS: reference sample}
\label{sec:vipers}

The VIMOS Public Extragalactic Redshift Survey
(VIPERS\footnote{http://vipers.inaf.it}) \citep{vipers_guzzo_2014} is
an on going spectroscopic survey whose aim is to map the detailed
spatial distribution of galaxies. The survey is made of two distinct
fields inside the CFHTLS W1 and W4 fields. The total survey area is
$24 \ \text{deg}^2$. VIPERS spectra are the results of $440$h of
observation at the Very Large Telescope (VLT) in Chile. Galaxies were
selected to have $z>0.4$ using the following colour criteria:
\begin{equation}
(r-i) > 0.5 (u-g) \ \ \text{OR} \ \ (r-i) > 0.7  \ .
\end{equation}

\begin{figure}
\begin{center}
\includegraphics[scale=0.35]{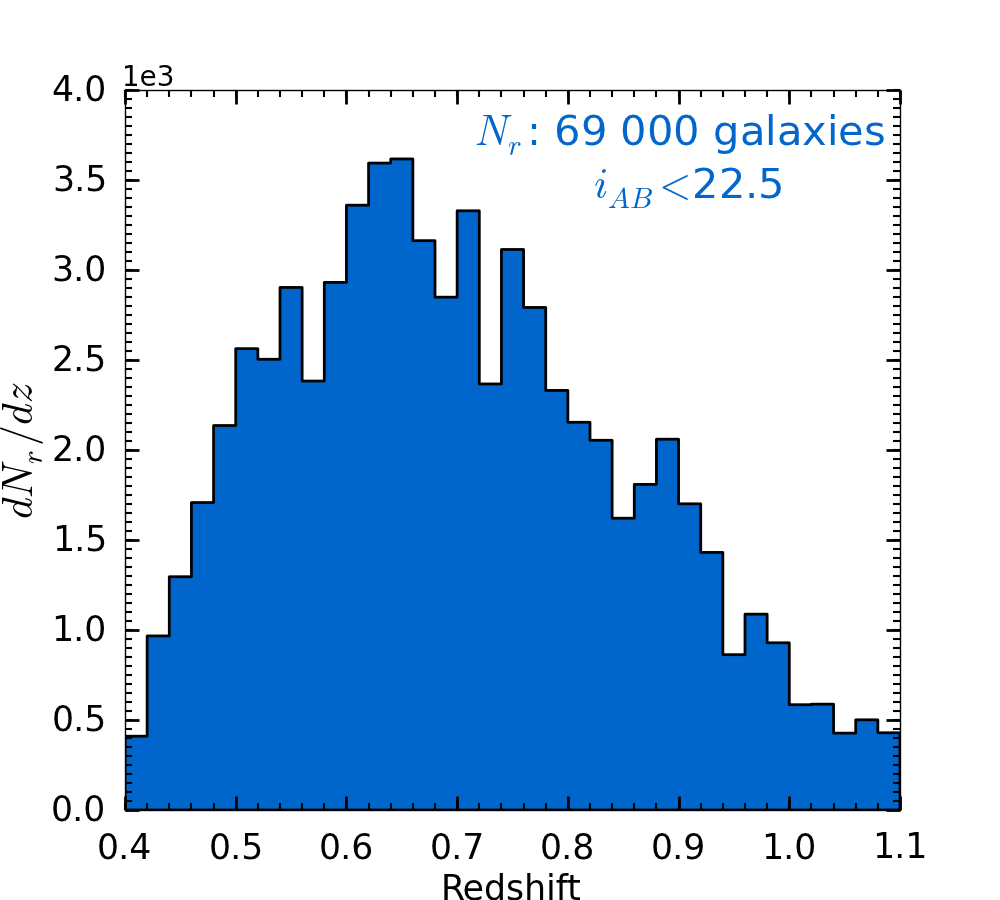} 
\end{center}
\caption{The redshift distribution of the reference sample built from VIPERS data with $i_{\text{AB}}<22.5$ and assuming a bin width $\delta z= 0.02$.}
\label{fig_nz_ref_sample}
\end{figure}

\noindent The 1 $\sigma$ random error in the measured VIPERS redshift is: $\sigma_z= 0.00047(1+z)$.

Our reference sample is made from a selection of VIPERS objects in two
separate fields, W1 and W4, outside CFHTLS masks and with secure
spectroscopic redshifts ($CL \ > \ 95\%$) corresponding to flags:
2,3,4 and 9 inside the redshift range $[0.4,1.1]$. The resulting
reference sample is composed of $\text{N}_{\text{r}} \sim 69 \ 000$
galaxies with $i_{\text{AB}}<22.5$ over an area of
$\sim 24 \ \text{deg}^2$. It corresponds to the reference population
used in all the analysis presented in this paper. Its redshift
distribution is shown in Figure~\ref{fig_nz_ref_sample}.

\subsubsection{CFHTLS: unknown sample}
\label{sec:cfhtls}

The Canada-France-Hawaii Telescope Legacy Survey
(CFHTLS\footnote{http://www.cfht.hawaii.edu/Science/CFHLS/}) Wide
includes four fields labelled W1, W2, W3 and W4. Complete
documentation of the CFHTLS-T0007 release can be found at the
CFHT\footnote{http://www.cfht.hawaii.edu/Science/CFHLS/T0007/}
site. In summary, the CFHTLS-Wide is a five-band survey of
intermediate depth. It consists of $171$ MegaCam deep pointings (of
$1 \ \text{deg}^2$ each) which, as a consequence of overlaps, consists
of a total of $\sim 155 \ \text{deg}^2$ in four independent contiguous
patches, reaching a $80\%$ completeness limit in AB of $u^* = 25.2$,
$g = 25.5$, $r = 25.0$, $i = 24.8$, $z = 23.9$ for point sources.

In this work we focused on the W1 and W4 fields in common with VIPERS
and used the magnitudes from the VIPERS Multi-Lambda Survey
\citep{moutard_2016a} which is based on the CFHTLS-T0007
photometry. We selected all galaxies in the same region of the sky
covered by VIPERS and which are outside CFHTLS masks resulting in a
sample of $\sim 570 \ 000$ galaxies over $\sim 24$
$\text{deg}^2$. Since we use a sample of VIPERS galaxies in the
redshift range $0.4<z<1.1$ we will not be able to measure any signal
outside this interval. Unknown objects outside this range will bias
the overall redshift distribution since it is normalised to the total
number of unknown galaxies following Equation\eqref{normalisation}. To
reduce this problem we selected objects with a photometric redshift
matching the range $[0.5;1]$ in redshift. Considering the number of
photometric sources at the edges and the photometric redshift
accuracy, we can expect to have less than 1\% of objects outside the
redshift range covered by the reference sample. The resulting
population corresponds to the parent unknown sample. This parent
sample is then divided into two samples: a bright sample with
$i_{\text{AB}}<22.5$ chosen to match the magnitude limit of the
reference population from VIPERS; and a faint sample whose galaxies
have magnitudes $22.5<i_{\text{AB}}<23$. These are the samples for
which we recover the redshift distributions in this paper.

\begin{figure}
\begin{center}
\includegraphics[scale=0.35]{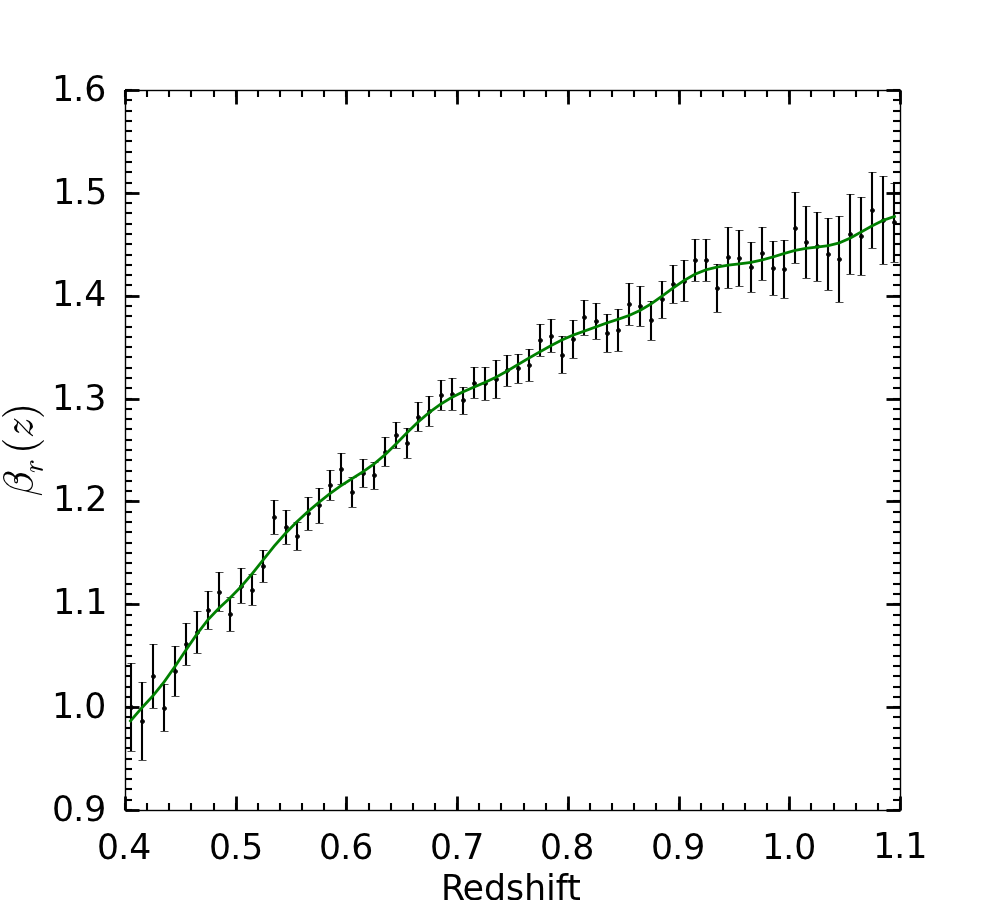} 
\end{center}
\caption{Clustering amplitude evolution of the reference sample normalised to $1$ at $z_{\text{0}}=0.4$. The solid line is the smoothed version used in this paper.}
\label{fig_cfht_vipers_clust_amplitude}
\end{figure}

\subsubsection{reference clustering amplitude measurement}
\label{sec:clust_ampl}

As previously seen in Section~\ref{sec:clusterz_formalism} the
determination of a clustering redshift distribution requires the
knowledge of the evolution with redshift of the clustering amplitude
of the reference population, $\beta_{\text{r}}(z)$. This
quantity can be directly measured using equation\eqref{clust_amp} and
is shown in Figure~\ref{fig_cfht_vipers_clust_amplitude}. We also show
a smoothed version obtained by convolving the binned measurements
with a Hann filter of width $\Delta z = 0.02$. Since we are only
interested in the relative variation of $\beta_{\text{r}}(z)$ \-- see
Equations~\eqref{n_u_z_propto_beta} \& \eqref{normalisation} \-- we
chose to normalise this quantity to unity at $z_{\text{0}}= 0.4$. This
figure shows an increase of $\sim 40\%$ of the clustering amplitude
between redshift 0.5 to 1.1 which is in agreement with the analysis
performed by \cite{marulli_2013}.

\section{Tomographic sampling:}
\label{sec:tomographic}

As seen in Section~\ref{sec:clusterz_formalism} reducing the
variation of $\beta_{\text{u}}(z)$ is a key point of this method. In
this section we aim at demonstrating our ability to measure the
redshift distribution. To reduce the variation of
$\beta_{\text{u}}(z)$ we choose to work with tomographic subsamples of
the unknown population. One can then consider:
$\deriv \beta_{\text{u}} / \deriv z = 0$ , for each of these
subsamples. The tomography is done by selecting objects using their
photometric redshifts based on the marginalization over the redshift
of all the models ($Z_{\text{ML}}$ in Lephare).

\subsection{Photometric redshifts estimation}
\label{sec:photoz_estimation} 

The photometric redshifts used in this paper come from the VIPERS-MLS
and are described in \cite{moutard_2016a}. The photometry combines
optical data from the CFHTLS-T0007 with near-infrared data (limited at
Ks$_{\text{AB}} < 22$). The authors have used ISO magnitudes that
provide the best estimate of galaxy colour and corrected them for a
mean difference between ISO and AUTO magnitudes (over the g, r, i and
Ks bands). This was done in order to recover a good approximation of
the galaxy total flux while keeping the best determination of the
galaxy colours. In our case this recalibration is important since it
leads to a smoother surface density fluctuation from tile to tile.

\FloatBarrier
\begin{figure}
\begin{center}
\includegraphics[scale=0.34]{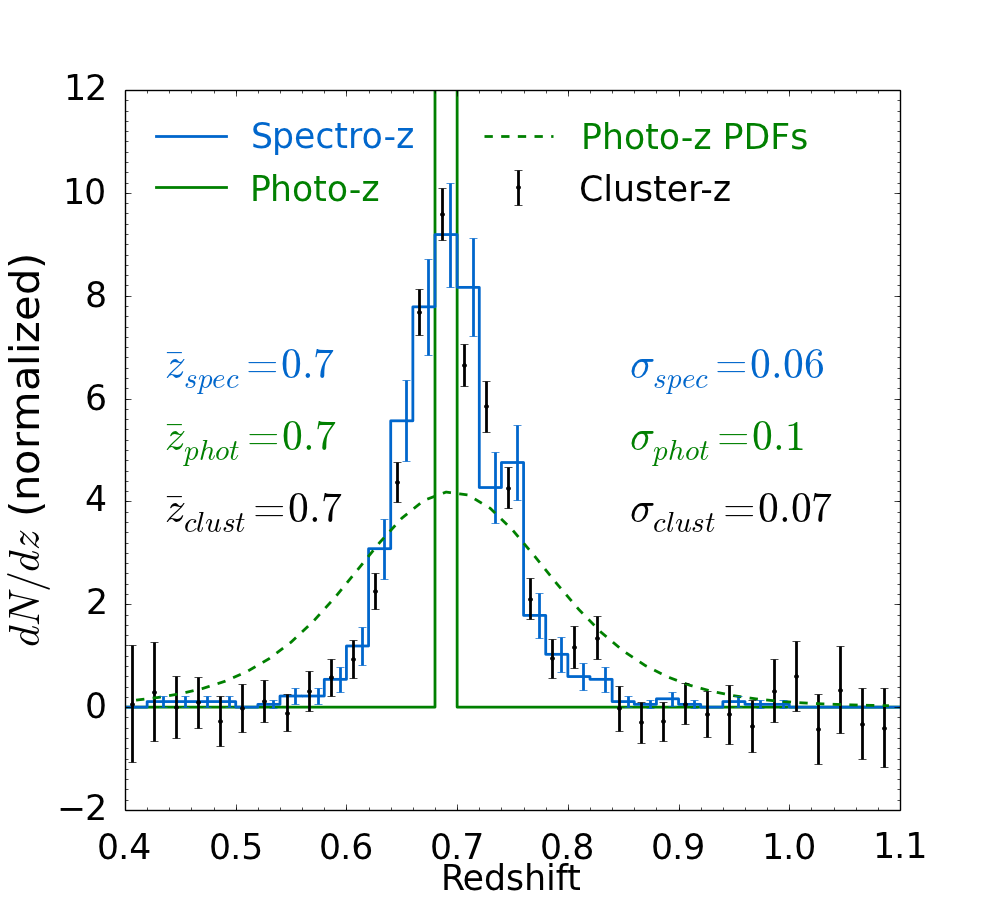} 
\end{center}
\caption{Example of the cluster-z distribution (black) obtained from Equation~\ref{n_u_z_propto_beta} for a tomographic sample selected using the photo-z (green line). The dashed green line shows the redshift distribution obtained when summing the photo-z PDFs. The blue line shows the spectroscopic redshift distribution with Poisson error bar of the VIPERS sources selected using their photometric redshifts to match the tomographic bin.} 
\label{fig_tomo_i225_thinbin}
\end{figure}

\begin{figure*}
\begin{center}
\includegraphics[scale=0.6]{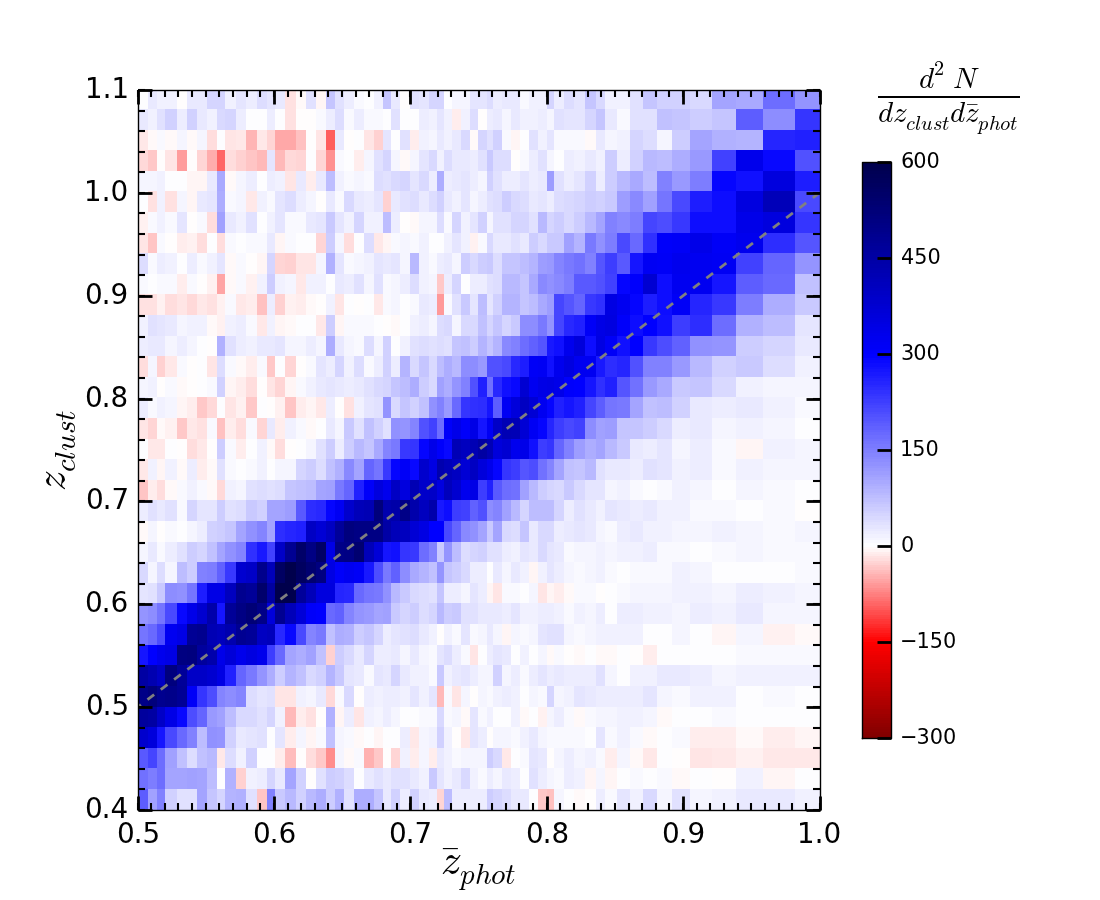}
\caption{Density map showing the $2 \ 380$ clustering measurements made in Section~\ref{sec:same_mag_tomo}. Each vertical line corresponds to a clustering redshift distribution measured in a tomographic sample of mean photometric redshift $\bar{z}_{\text{phot}}$. Each column is normalised to match the number of unknown objects.}
\label{fig_same_mag_tomo_zcl_zph}
\end{center}
\end{figure*} 

\subsection{Magnitude limit for both samples: $i<22.5$}
\label{sec:same_mag_tomo}

We selected objects with $i_{\text{AB}}<22.5$ in the unknown
population. The resulting sample contains
$\text{N}_{\text{u}} \sim 203 \ 000$ galaxies.

We split them into $68$ tomographic subsamples of $3 \ 000$ objects
each. Thus, we measure the integrated cross-correlation from few kpcs
to several Mpcs in reference slices of width
$\delta z= 0.02$.

Figure~\ref{fig_tomo_i225_thinbin} shows the recovered clustering
redshift distribution for a particular tomographic bin selected using
the photometric redshift. We also show the redshift distribution
obtained when using photo-z PDFs. This PDF is obtained by stacking
individual PDFs defined as a gaussian:
$G(z_{\text{phot}},\sigma= z_{\text{phot,max}}-z_{\text{phot,min}})$,
where $z_{\text{phot,min/max}}$ are the 1 $\sigma$ lower/upper limit,
respectively. This plot shows the ability of reconstructing the
redshift distribution with the clustering method.  Recovered
distributions (black dots) is significantly narrower that the photo-z
PDF (dashed green) and consistent with the distribution of
spectroscopic VIPERS galaxies (in blue) selected on their photometric
redshift to match the selected tomographic bin.

Note that this is not a rigorous comparison since the spectroscopic
sources show in blue are not exactly the same objects considered in
the unknown sample. Moreover, since there are only few
objects in this distribution one can only compare the statistical
properties which are expected to be similar. All distributions are
normalised to unity.

In the same way, we measured the clustering redshifts distributions
for all the 68 tomographic subsamples. The results of these
$68 \times 35 = 2 \ 380$ measurements of $\bar{\omega}_{\text{ur}}$
are translated into redshift distributions following
equation\eqref{n_u_z_propto_beta}.

In Figure~\ref{fig_same_mag_tomo_zcl_zph} each vertical line
corresponds to a clustering redshift distribution measured in a
tomographic sample of mean photometric redshift
$\bar{z}_{\text{phot}}$ similar to the one shown in
Figure~\ref{fig_tomo_i225_thinbin}. Figure~\ref{fig_same_mag_tomo_zcl_zph}
shows the corresponding redshift distributions in the
$\left( z_{\text{clust}};z_{\text{phot}} \right)$ plane and
illustrates the global agreement between cluster and photo-z. Negative
values correspond to stochastic density fluctuations and are not
statistically significant.

To compare these two measurements in a more quantitative way we
compute the accuracy of the estimate of the mean redshift of a
distribution as:
$\sigma = \sigma_{\Delta \bar{z}} / (1 + \bar{z}_{\text{spec}})$,
where
$\Delta \bar{z} = | \bar{z}_{\text{clust/phot}} -
\bar{z}_{\text{spec}} | $
is the difference between the mean clustering redshift and the mean
spectroscopic redshift of a distribution.

\begin{figure}
\begin{center}
\includegraphics[scale=0.35]{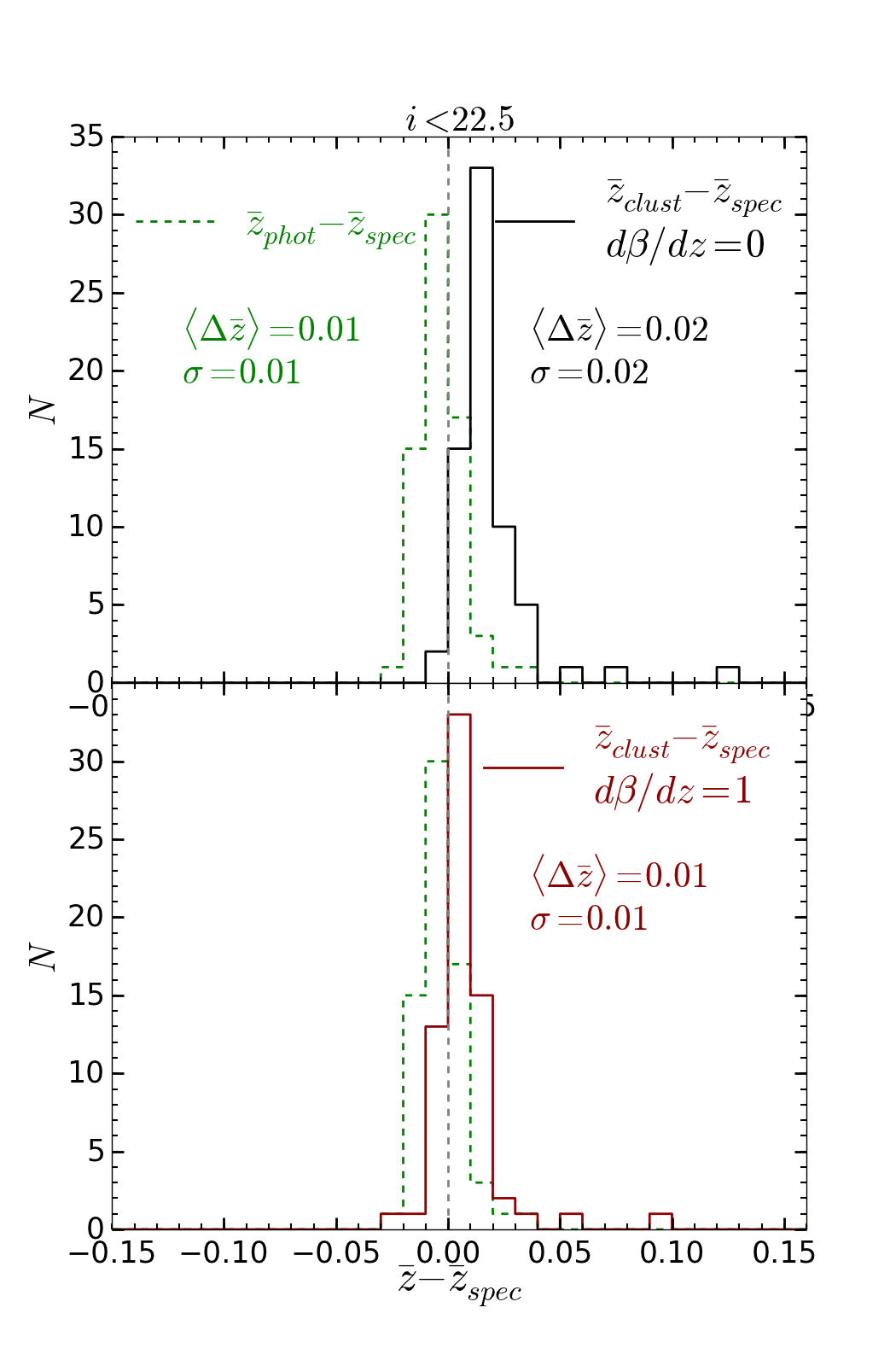} 
\end{center}
\caption{\underline{TOP PANEL:} Histograms showing the distribution of the difference between the mean of the clustering/photometric redshift distribution and the mean of the spectroscopic redshift distribution: $ \bar{z}_{\text{clust/phot}} - \bar{z}_{\text{spec}} $ ( black and dashed green lines, respectively). Cluster-z measurements were made considering $\deriv \beta_{\text{u}} / \deriv z =0$. \newline
 \underline{BOTTOM PANEL:} Same quantities as in the top panel but the cluster-z measurements were performed considering a linear evoluton for the clustering amplitude of the unknown population: $\deriv \beta_{\text{u}} / \deriv z =1$.}
\label{fig_same_mag_tomo_residues}
\end{figure}

We use the normalised median absolute deviation to estimate the accuracy as previously defined:
\begin{equation}
\sigma_{\Delta \bar{z}}= 1.48 \times \text{median}( | \bar{z}_{\text{clust/phot}} - \bar{z}_{\text{spec}} |) \ ,
\end{equation}
where the mean redshifts are computed as:
\begin{equation}
\bar{z} = \frac{1}{\sum \deriv \text{N} / \deriv z} \left( \sum_i z_i \frac{\deriv \text{N}_i}{\deriv z} \right) \ .
\end{equation}

\newpage

\noindent For each cluster-z distribution we show on the top
panel of Figure~\ref{fig_same_mag_tomo_residues} the difference
$ \bar{z}_{\text{clust/phot}} - \bar{z}_{\text{spec}} $. We see that
cluster-z and photo-z are in relatively good agreement. This figure
demonstrates the ability of cluster-z to infer redshift distributions
of a sample for which photometric redshifts are known and can be used
to reduce the variation of $\beta_{\text{u}}(z)$ by selected
subsamples localised in redshift.

The lower panel shows the $ \bar{z}_{\text{clust/phot}} - \bar{z}_{\text{spec}} $ residuals when
considering a linear evolution of the unknown clustering amplitude
$\deriv \beta_{\text{u}} / \deriv z = 1$ instead of a constant
evolution following \cite{rahman_2015}. This tomographic sampling
approach does not allow us to estimate $\beta_u(z)$ using photo-z due
to the thickness of the selected bins. This will be done in the colour
sampling approach in Section~\ref{sec:same_mag_colors_beta_u}.

We remind the reader that in this analysis the photo-z information is
only used to select subsamples localised in redshift in a
preprocessing step. The only goal of photo-z here is to provide an
easy way to select redshift distributions narrow in redshift but one
can use any other way to do so. Once these subsamples are built the
only used information is the over/under-density around
reference galaxies is used to estimate the redshift
distribution. Then cluster-z and photo-z methods could be used
separately for validation and/or combined together.

\subsection{Fainter unknown sample: $22.5<i<23$}
\label{sec:fainter_tomo}

\noindent This section shows our ability to measure clustering redshifts when the unknown sample is fainter than the reference one.

\begin{figure}
\begin{center}
\includegraphics[scale=0.37]{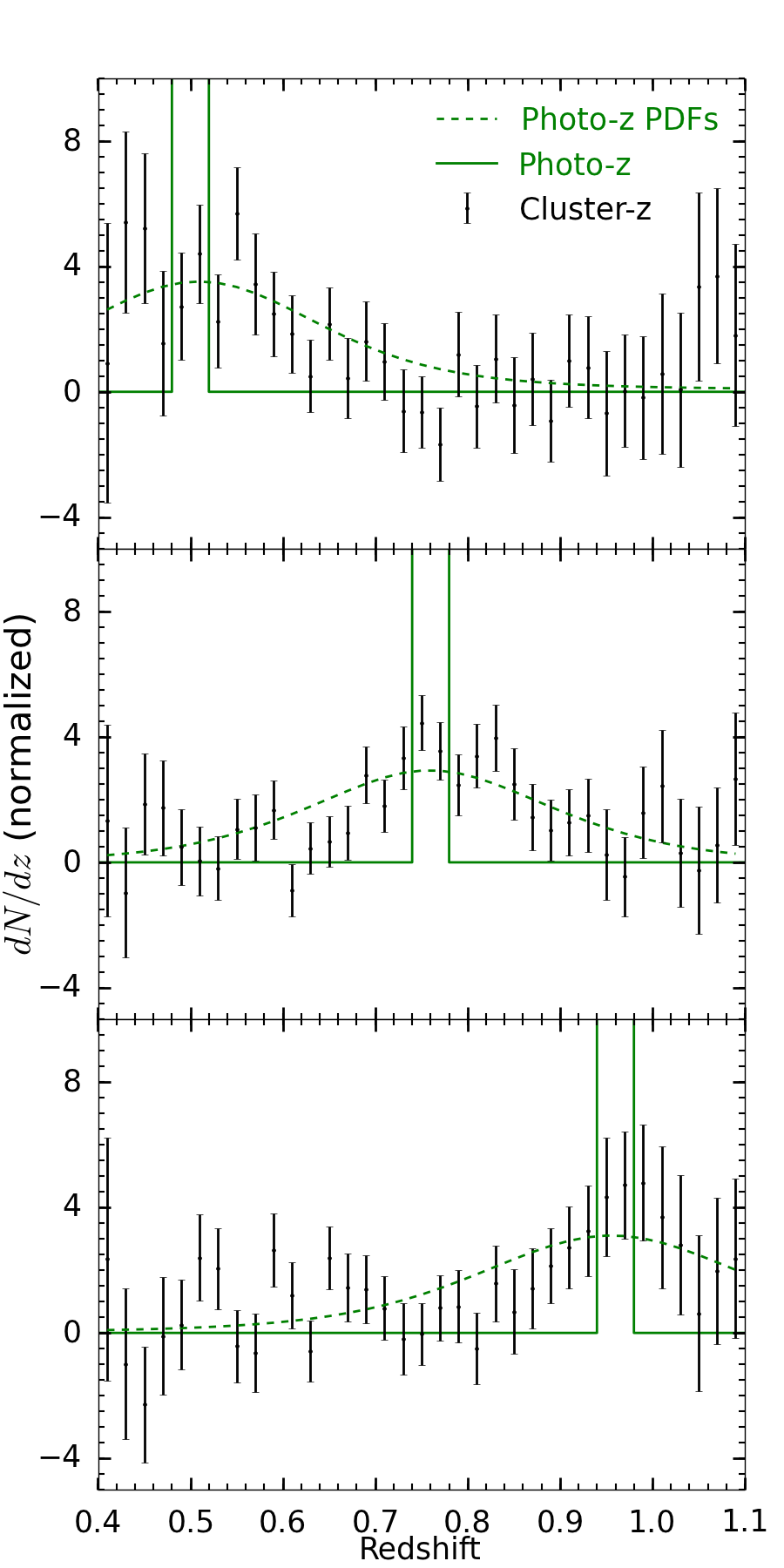}
\end{center}
\caption{The cluster-z and photo-z distributions for several tomographic bins at magnitude $22.5<i<23$. Cluster-z (black points) are in agreement with photo-z PDFs (green dashed line) demonstrating that this method is able to extract the desired signal. The results for other bins are available online.}
\label{fig_fainter_mag_tomo_all_binsparts_1_3}
\end{figure}

\noindent Since we are not looking at the spectral energy distribution (SED) but at the clustering of objects and since all objects cluster with each other \-- regardless of their magnitude \-- we expect a signal \citep{menard_2013,rahman_2015_IR,rahman_2015_sdssphotz}.

Nevertheless, since at a given redshift fainter objects are less
massive we can expect a lower signal than in the previous case. To
illustrate this we use the same reference population used
previously with $i_{\text{ref}} < 22.5$ but we select objects from
the unknown sample with $22.5 < i_{\text{unk}} < 23$. This
leads to an unknown faint sample made of
$\text{N}_{\text{u}} = 88 \ 000$ galaxies. To be coherent with the
previous case we build tomographic subsamples of
$\text{N}_{\text{u}} = 3 \ 000$ galaxies.

~~\\

\noindent The resulting clustering-based redshifts distributions for 3
selected bins that span the all redshift range are shown in black in
Figure~\ref{fig_fainter_mag_tomo_all_binsparts_1_3}. The measurements
of all bins are available online. By computing the quantity
$\bar{z}_{\text{clust}} - \bar{z}_{\text{phot}}$ for each distribution
one can estimate $\left\langle \Delta \bar{z} \right\rangle$ and the
$\sigma$ and then compare cluster-z to photo-z, see
Table~\ref{table_1}.

\newpage

\noindent One can see that clustering-redshift distributions are in agreement with photo-z PDFs. Indeed we find $\left\langle \Delta \bar{z} \right\rangle=0.05$ and $0.04$ \& $\sigma= 0.06$ when considering $\deriv \beta_{\text{u}} / \deriv z =0$ and $\deriv \beta_{\text{u}} / \deriv z =1$, respectively. This demonstrates that the signal could be detected even when the reference and unknown populations do not have the same magnitude limit.

\newpage

\begin{table}
\begin{tabular}{lll}
\hline
 \multicolumn{3}{|c|}{$22.5<i<23$} \\
	\hline
\multicolumn{3}{|c|}{$\bar{z}_{\text{clust}}  - \bar{z}_{\text{phot}}$} \\
\hline	
 \multicolumn{1}{|c|}{}   & \multicolumn{1}{c|}{$\deriv \beta_{\text{u}} / \deriv z =0$} & \multicolumn{1}{c|}{$\deriv \beta_{\text{u}} / \deriv z =1$} \\
    \hline
\multicolumn{1}{|c|}{$\left\langle  \Delta \bar{z} \right\rangle$}    & \multicolumn{1}{c|}{0.05} & \multicolumn{1}{c|}{0.04} \\
\multicolumn{1}{|c|}{ $\sigma$} &  \multicolumn{1}{c|}{0.06}  & \multicolumn{1}{c|}{0.06} \\
	\hline
\end{tabular}
\caption{Comparison between the mean clustering redshift and the mean photometric redshift from the distributions. This comparison is done when considering $\deriv \beta_{\text{u}} / \deriv z =0$ and $\deriv \beta_{\text{u}} / \deriv z =1$. In both cases the two methods are in agreement. }
\label{table_1}
\end{table}

\noindent In the context of large imaging experiments the
requirements on spectroscopic redshifts are challenging. In
particular it is difficult to make complete spectroscopic samples
down to magnitudes $i_{\text{AB}}=24$ which is the magnitude limit of
large imaging surveys like Euclid. This property of clustering
redshift is therefore of great interest.

\section{Colour sampling}
\label{sec:colors_sampling}

In this section we aim at freeing the clustering-based redshift
estimation technique from the need of photometric redshifts to
preselect subsamples localised in redshift and quantify the resulting
accuracy.

\begin{figure*} 
\begin{center}
\includegraphics[scale=0.4, angle =90]{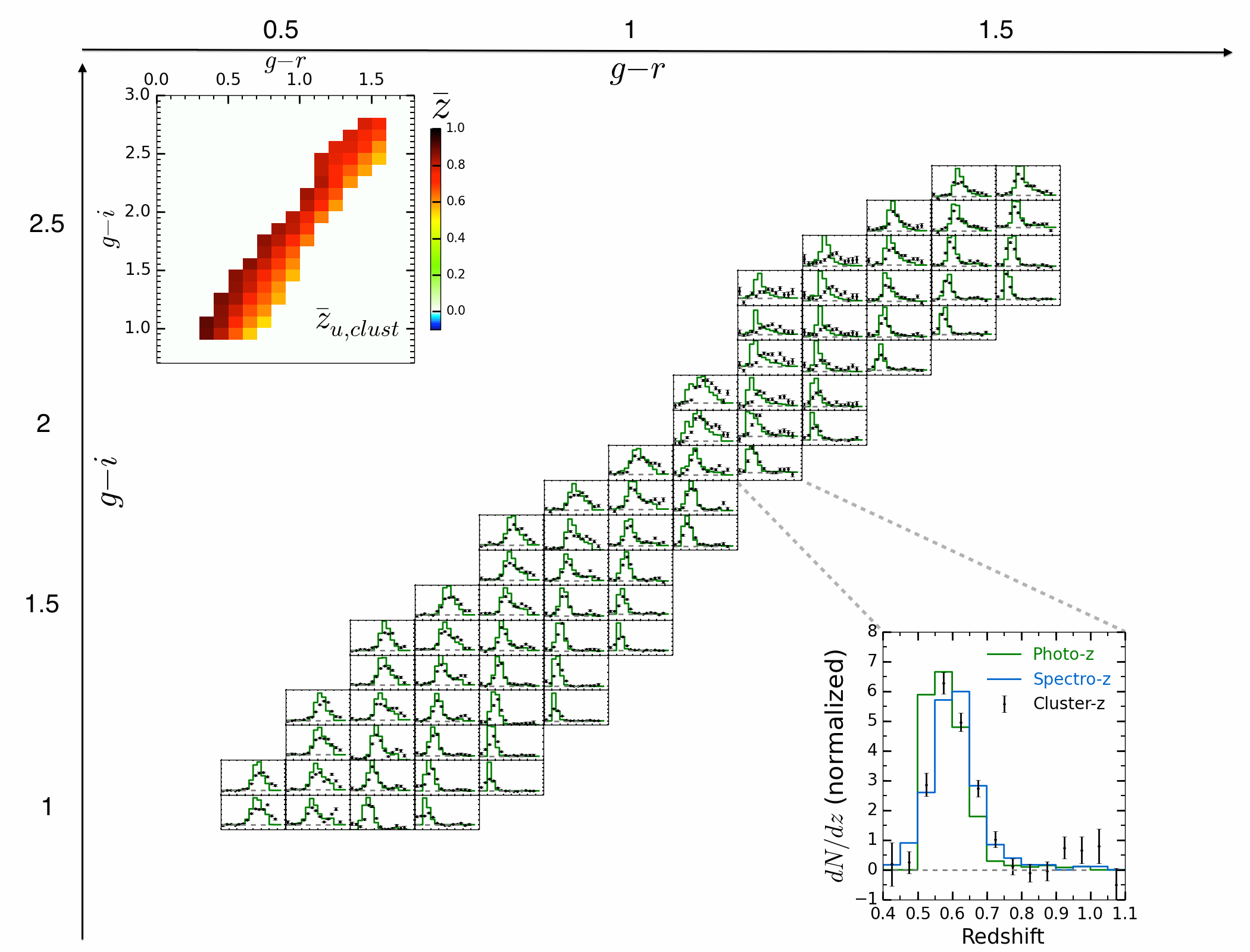} 
\end{center}
\caption{Cluster-z (black points) and photo-z distributions in each cell of the all colour-space for $i_{\text{unk}}<22.5$. The top left panel shows the evolution of the mean redshift with colours. A zoom for a given cell is shown in the bottom right panel.}
\label{fig_huge_map_i225}
\end{figure*}

\subsection{Magnitude limit for both samples: $i<22.5$}
\label{sec:same_mag_colors}

First we look at an unknown population with the same
limiting magnitude of the reference sample. In this case we
expect the reference sample to be a representative sample of
our unknown population. Then, the colour-redshift relation of
both samples should be the same on average.

To reduce the effect of the clustering amplitude evolution with
redshift of the unknown sample, $\beta_{\text{u}}(z)$, we
build subsamples in colour-space. Working on the $(g-i;g-r)$ plane, we
choose a binning size of $\Delta_{g-r/i} = 0.1$. By construction the
redshift distribution in each of these cells will be narrower than the
one of the initial sample.

Then we measure the clustering redshift distribution in each cell. All
these distributions across the colour space as well as their
corresponding photometric and spectroscopic redshift distributions can
be seen in Figure~\ref{fig_huge_map_i225}. The central part of this
plot shows the redshift distribution evolution with colours.

\begin{figure*}
\begin{center}
\includegraphics[scale=0.39]{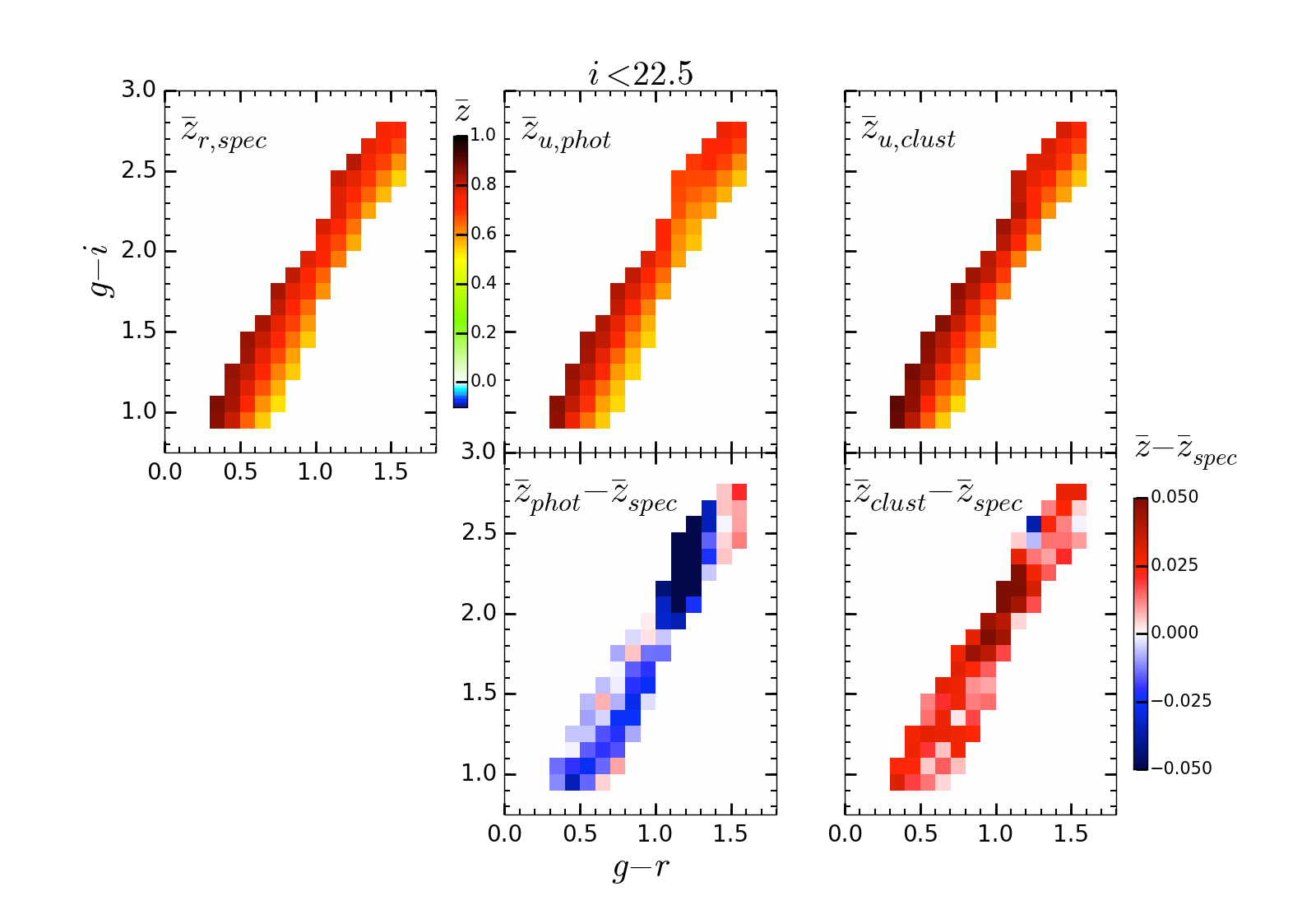}   
\end{center}
\caption{\underline{TOP PANEL:} Mean redshift evolution through colour-space for both reference and unknown samples. \newline
 \underline{BOTTOM PANEL:} Difference between the estimated mean redshift \-- cluster or photo-z  \-- and the true mean redshift from spectroscopic measurements. One can notice a bad region around $g-r=1.2$ for photo-z. This points out a systematic effect in the photo-z measurements.}
\label{fig_same_mag_colors_maps}
\end{figure*}

When $g-r$ decreases the redshift increases. This corresponds to the 4
000 \AA \ break going through the $r$-band between redshift 0.4 and
1. On the contrary $g-i$ increases with redshift. This is due to the
4 000 \AA \ break approaching the $i$-band. The bottom right panel is
a zoom in one cell where one can see the clustering redshift
distribution in black and the photo-z distribution in green. The top
left panel shows with a colour code the evolution of the mean
clustering redshift with colours. This map gives a direct view of the
colour-redshift relation reconstructed by clustering redshifts.

\newpage

\noindent In the same way the upper panels in
Figure~\ref{fig_same_mag_colors_maps} show the mean redshift evolution
through colour space but for both reference and
unknown samples. This figure also shows in the bottom panels
the differences: $\bar{z}_{\text{cl}} - \bar{z}_{\text{spec}}$ and
$\bar{z}_{\text{ph}} - \bar{z}_{\text{spec}}$. One can notice the
large residuals for photo-z at $(g-i, g-r) \sim (2.2,1.2)$. They
reveal the presence of a systematic effect affecting the photo-z
estimate. One can note that the template library has been calibrated
with the CFHTLenS optical photometry whose absolute calibration
differs by $\sim 0.15$ mag in the $z$-band in comparison with the
T0007 one \citep{moutard_2016b}. This could explain part of the
photo-z bias that is observed for red galaxies. We leave to a future
work a more detailed exploration of this effect.

To compare, in a more quantitative way, the ability of
cluster-z to reproduce the colour-redshift relation compared
to photo-z we compte the residual
$\bar{z}_{\text{cl/ph}} - \bar{z}_{\text{spec}}$ and summarize the
results in Table~\ref{table_2}. We find
$\left\langle \Delta \bar{z} \right\rangle = 0.02$ for cluster-z and
photo-z while they have $\sigma= 0.02$ and $0.03$, respectively. This
shows that in the colour sampling approach, cluster-z and photo-z have
similar accuracy with respect to spectro-z. Nevertheless one can note
that here we use only 3 bands to extract subsamples from the
unknown population of objects. The resulting cluster-z are
compared to photo-z while photo-z were obtained by combining optical
and near-infrared data. This is encouraging because there is still
plenty of information to be added.  Other galaxy properties such as
size, brightness and ellipticity can be used in addition to the colours
to improve the cluster-z estimation.

\begin{table}
\begin{tabular}{lll}
\hline
\multicolumn{3}{|c|}{$i<22.5$} \\
\hline
\multicolumn{1}{|c}{}    & \multicolumn{1}{c|}{$\bar{z}_{\text{phot}}  - \bar{z}_{\text{spec}}$} & \multicolumn{1}{c|}{$\bar{z}_{\text{clust}}  - \bar{z}_{\text{spec}}$} \\
    \hline
     \multicolumn{2}{|c|}{-}	&  \multicolumn{1}{c|}{$\deriv \beta_{\text{u}} / \deriv z =0$}  \\
	\hline
 \multicolumn{1}{|c|}{$\left\langle  \Delta \bar{z} \right\rangle$} &  \multicolumn{1}{c|}{0.02} &  \multicolumn{1}{c|}{0.02} \\
  \multicolumn{1}{|c|}{$\sigma$} &   \multicolumn{1}{c|}{0.04}  &  \multicolumn{1}{c|}{0.03} \\
	\hline
\end{tabular}
\caption{Comparison between the mean spectro/photo/cluster-z when considering $\deriv \beta_{\text{u}} / \deriv z =0$. The bais and scatter of these two approaches are similar when comparing to spectroscopic redshift.  }
\label{table_2}
\end{table}

\subsection{Evolution of the unknown clustering amplitude $\beta_{\text{u}}(z)$}
\label{sec:same_mag_colors_beta_u}

In this section we investigate the validity of the assumption made on the evolution of the clustering amplitude of the unknown sample, $\beta_{\text{u}}(z)$.

Since we know the photometric redshifts for the unknown population we can use them to estimate the true evolution with redshift of the clustering amplitude, $\beta_{\text{u}}(z)$. To do so, we apply the same procedure used in the measurement of the reference sample clustering amplitude, $\beta_{\text{r}}(z)$ ( see section~\ref{sec:clust_ampl}) following equation~\eqref{clust_amp}. This procedure is applied in each cell of the colour-space. Results are shown in Figure~\ref{fig_same_mag_colors_beta_u} where we report the measured $\beta_{\text{u}}(z)$ based on photometric redshifts (in black) whereas in Figure~\ref{fig_cfht_vipers_clust_amplitude} and in the equations of Section~\ref{sec:clusterz_formalism} $\beta$ is a function of $z_{\text{spec}} \sim z_{\text{true}}$. One can note that in these regions of the colour-space the clustering amplitude $\beta_{\text{u}}(z)$ seems to evolve linearly with redshift. For this reason we also show a linear fit (in green).

\begin{figure*}
\begin{center}
\includegraphics[scale=0.3]{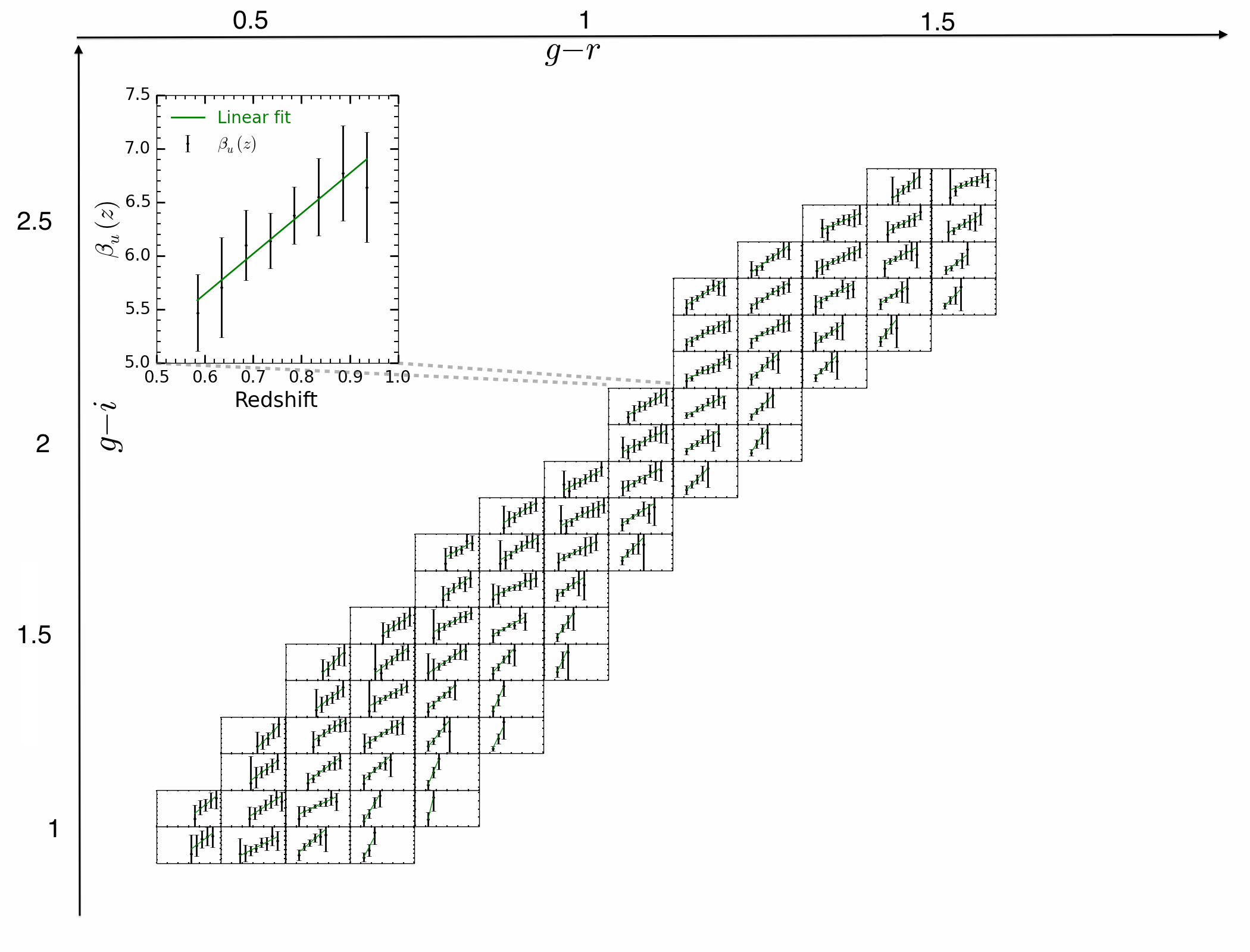}   
\end{center}
\caption{Black points show the measured clustering amplitude of the unknown sample $\beta_{\text{u,true}}(z)$ for each cell in the colour-space. The green line corresponds to a linear fit. This is computed using photo-z. A zoom for a given cell is shown in the top right panel.}
\label{fig_same_mag_colors_beta_u}
\end{figure*}
\FloatBarrier

\noindent Based on these measurements we can then compute the offset in the mean redshift, $\epsilon \equiv \bar{z}_{\text{estimated}} - \bar{z}_{\text{true}}$, due to the non-evolution hypothesis we made on the unknown clustering amplitude: $\deriv \beta_{\text{u}} / \deriv z = 0$. The histogram showing the resulting offsets in the mean redshift estimates is visible in Figure~\ref{fig_same_mag_colors_beta_u_offset}. In this case the effect of considering $\deriv \beta_{\text{u}} / \deriv z = 0$ is a bias of order $0.02$ in the mean redshift estimate.

\begin{figure}
\begin{center}
\includegraphics[scale=0.3]{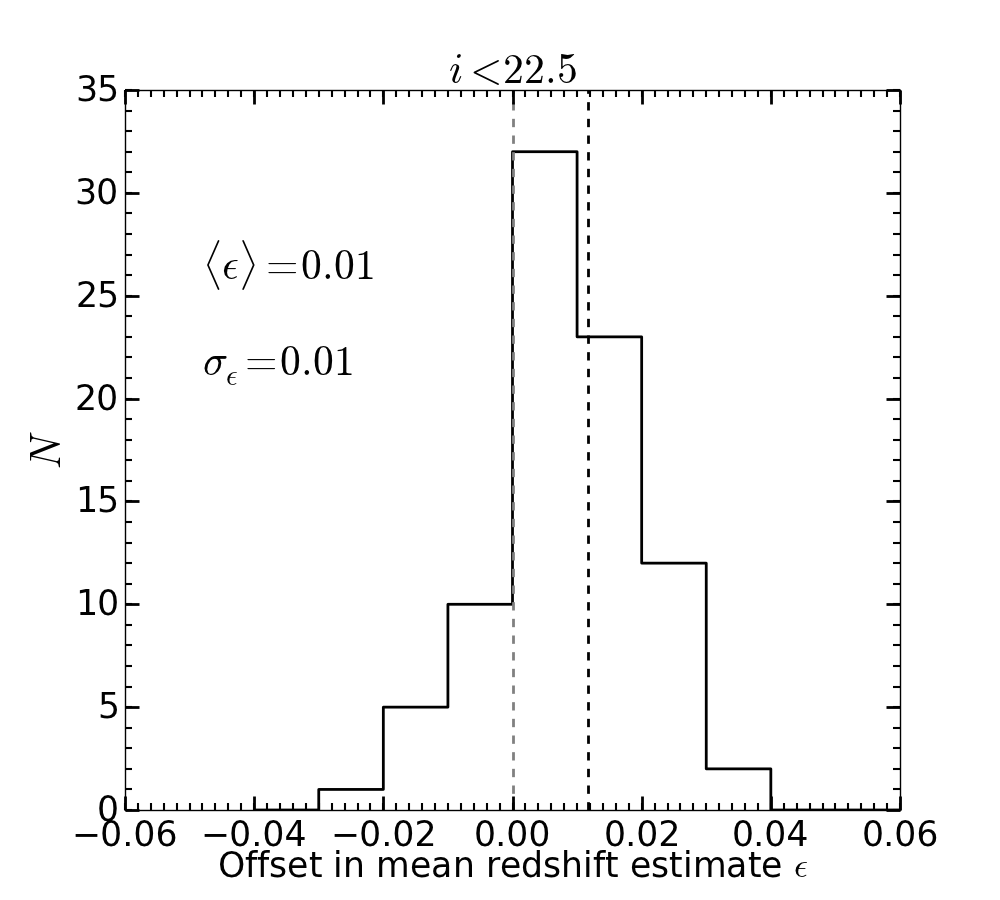}  
\end{center}
\caption{Offset in the mean redshift estimates due to the evolution of the unknown clustering amplitude $\beta_{\text{u}}(z)$. Considering no evolution for $\beta_{\text{u}}$ leads to a bias of 0.02 in the mean redshift recovered by the clustering-based redshift estimation method.}
\label{fig_same_mag_colors_beta_u_offset}
\end{figure}

Moreover since the clustering amplitudes we measured seem to be linear in redshift we can estimate the cluster-z distributions obtained in Section~\ref{sec:same_mag_colors} when considering $\deriv \beta_{\text{u}} / \deriv z = 1$ in Equation~\ref{n_u_z_propto_beta} instead of $\deriv \beta_{\text{u}} / \deriv z= 0$. The results of these new measurements are summarised in Table~\ref{table_3}.

\begin{table}
\begin{tabular}{llll}
\hline
 \multicolumn{4}{|c|}{$i<22.5$} \\
 \hline
 \multicolumn{1}{|c}{}     &  \multicolumn{1}{c|}{$\bar{z}_{\text{phot}}  - \bar{z}_{\text{spec}}$} &  \multicolumn{2}{c|}{$\bar{z}_{\text{clust}}  - \bar{z}_{\text{spec}}$} \\
    \hline
\multicolumn{2}{|c|}{-} &  \multicolumn{1}{c|}{$\deriv \beta_{\text{u}} / \deriv z =0$} & \multicolumn{1}{>{\columncolor{Gray}}c|}{$\deriv \beta_{\text{u}} / \deriv z =1$}  \\
	\hline

\multicolumn{1}{|c|}{$\left\langle  \Delta \bar{z} \right\rangle$}    & \multicolumn{1}{c|}{0.02} &  \multicolumn{1}{c|}{0.02} & \multicolumn{1}{>{\columncolor{Gray}}c|}{0.01} \\
 \multicolumn{1}{|c|}{$\sigma$} &  \multicolumn{1}{c|}{0.04}  &  \multicolumn{1}{c|}{0.03} & \multicolumn{1}{>{\columncolor{Gray}}c|}{0.02} \\
	\hline
\end{tabular}
\caption{Same table than Table~\ref{table_2} but we add in grey the result when considering $\deriv \beta_{\text{u}} / \deriv z =1$. This sligthly improves the clustering redshift measurements.}
\label{table_3}
\end{table}

\begin{table*}
\begin{tabular}{lllllll}
\hline
  \multicolumn{4}{|c|}{$i<22.5$}  & \multicolumn{3}{>{\columncolor{Gray}}c|}{$22.5<i<23$} \\
 \hline
 \multicolumn{1}{|c}{}    & \multicolumn{1}{c|}{$\bar{z}_{\text{phot}}  - \bar{z}_{\text{spec}}$} & \multicolumn{2}{c|}{$\bar{z}_{\text{clust}}  - \bar{z}_{\text{spec}}$} &  \multicolumn{1}{>{\columncolor{Gray}}c|}{$\bar{z}_{\text{phot}}  - \bar{z}_{\text{spec}}$} & \multicolumn{2}{>{\columncolor{Gray}}c|}{$\bar{z}_{\text{clust}}  - \bar{z}_{\text{spec}}$} \\
	\hline
 \multicolumn{2}{|c|}{-} &  \multicolumn{1}{c|}{$\deriv \beta_{\text{u}} / \deriv z =0$}  & \multicolumn{1}{c|}{$\deriv \beta_{\text{u}} / \deriv z =1$} &  \multicolumn{1}{>{\columncolor{Gray}}c|}{-}  &  \multicolumn{1}{>{\columncolor{Gray}}c|}{$\deriv \beta_{\text{u}} / \deriv z =0$}  & \multicolumn{1}{>{\columncolor{Gray}}c|}{$\deriv \beta_{\text{u}} / \deriv z =1$}  \\
    \hline
\multicolumn{1}{|c|}{$\left\langle  \Delta \bar{z} \right\rangle$  } & \multicolumn{1}{c|}{0.02} &  \multicolumn{1}{c|}{0.02} & \multicolumn{1}{c|}{0.01}  & \multicolumn{1}{>{\columncolor{Gray}}c|}{0.03} & \multicolumn{1}{>{\columncolor{Gray}}c|}{0.05} & \multicolumn{1}{>{\columncolor{Gray}}c|}{0.04} \\
\multicolumn{1}{|c|}{ $\sigma$ } &  \multicolumn{1}{c|}{0.04}  &  \multicolumn{1}{c|}{0.03} & \multicolumn{1}{c|}{0.02} &  \multicolumn{1}{>{\columncolor{Gray}}c|}{0.05}  & \multicolumn{1}{>{\columncolor{Gray}}c|}{0.07} & \multicolumn{1}{>{\columncolor{Gray}}c|}{0.06} \\
	\hline
\end{tabular}
\caption{Same table than Table~\ref{table_3} but we add in grey the results when considering $\deriv \beta_{\text{u}} / \deriv z =0$ and $\deriv \beta_{\text{u}} / \deriv z =1$ in the case where the unknown population is fainter than the reference sample.}
\label{table_4}
\end{table*}
\FloatBarrier

\newpage

\noindent Finally we combined all cluster-z measurements to derive the global redshift distribution in Figure~\ref{fig_same_mag_colors_sum_nz_lin_and_cst}. The top panel shows the two photo-z distributions as well as the global clustering redshift distributions when accounting or not for a linear evolution of $\beta_{\text{u}}(z)$. These distributions are obtained by summing the distributions from each cells in colour-space including cells located in the bad region of the photo-z map ( see Figure~\ref{fig_same_mag_colors_maps}). Considering a linear evolution for $\beta_{\text{u}}(z)$ allows to correct the small distortion of the distribution. As expected it appears that the no-evolution assumption tends to slightly underestimate the number of galaxies at low redshift and to slightly overestimate it at high redshift.

\FloatBarrier
\begin{figure}
\begin{center}
\includegraphics[scale=0.32]{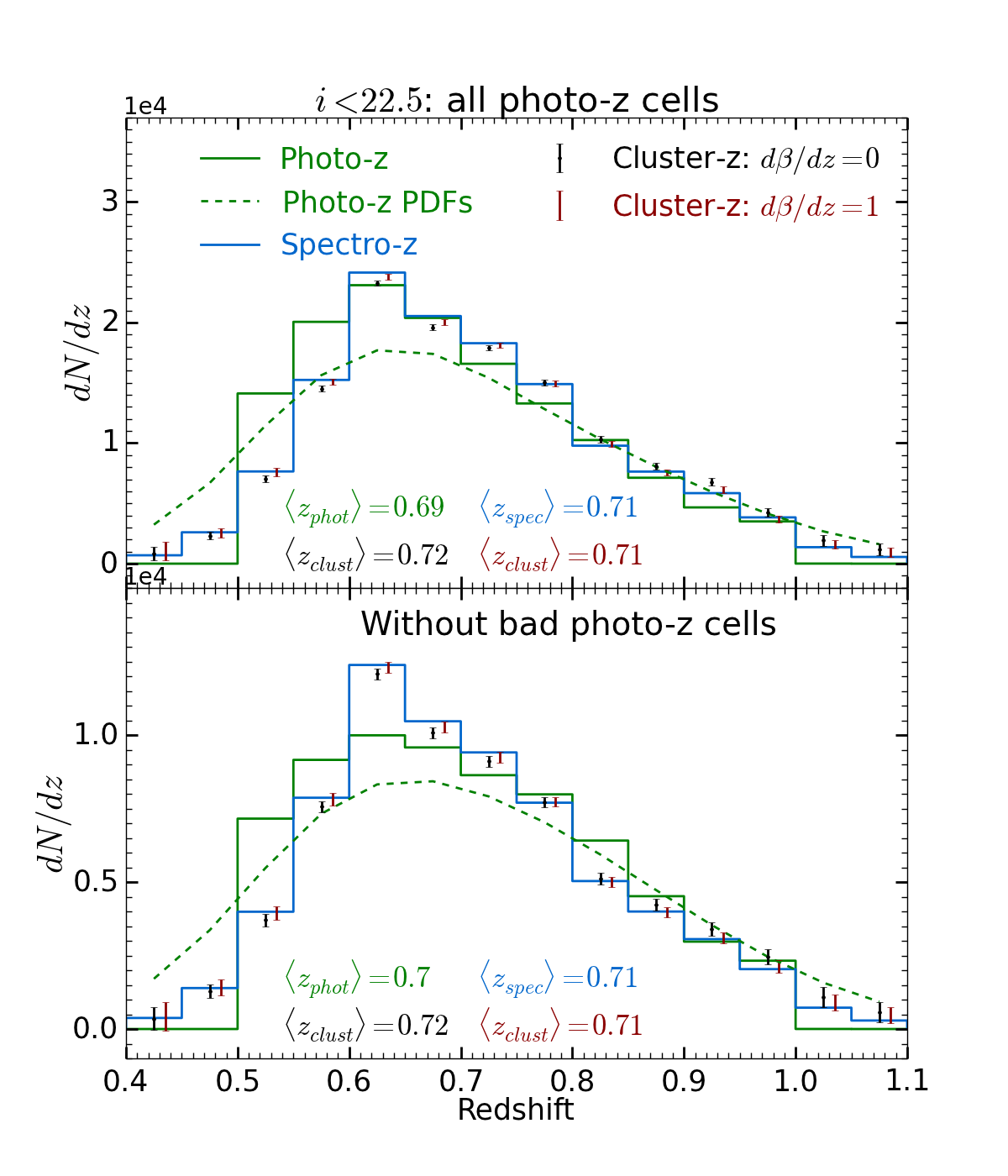}   
\end{center}
\caption{\underline{TOP PANEL:} global redshift distribution estimated by photo-z (green \& dashed green), spectro-z (blue) and by cluster-z (black dots) for $\deriv \beta_{\text{u}} / \deriv z = 0$. Red dots correspond to cluster-z with $\deriv \beta_{\text{u}} / \deriv z = 1$. These distributions are obtained by adding the distributions of the all cells of the colour-space, including the bad region visible in Figure~\ref{fig_same_mag_colors_maps}. \newline
\underline{BOTTOM PANEL:} As the top panel but excluding cells in the two columns around $g-r \sim 1.2$. }
\label{fig_same_mag_colors_sum_nz_lin_and_cst}
\end{figure}

The bottom panel of Figure~\ref{fig_same_mag_colors_sum_nz_lin_and_cst} shows the same quantities but when summing the distributions only from the "good" cells in colour-space. In this case we excluded cells located in the bad region of the photo-z map i.e cell in the two columns at $g-r \sim 1.2$.

\subsection{Fainter unknown sample: $22.5<i<23$}
\label{sec:fainter_colors}

\begin{figure*} 
\begin{center}
\includegraphics[scale=0.4, angle =90]{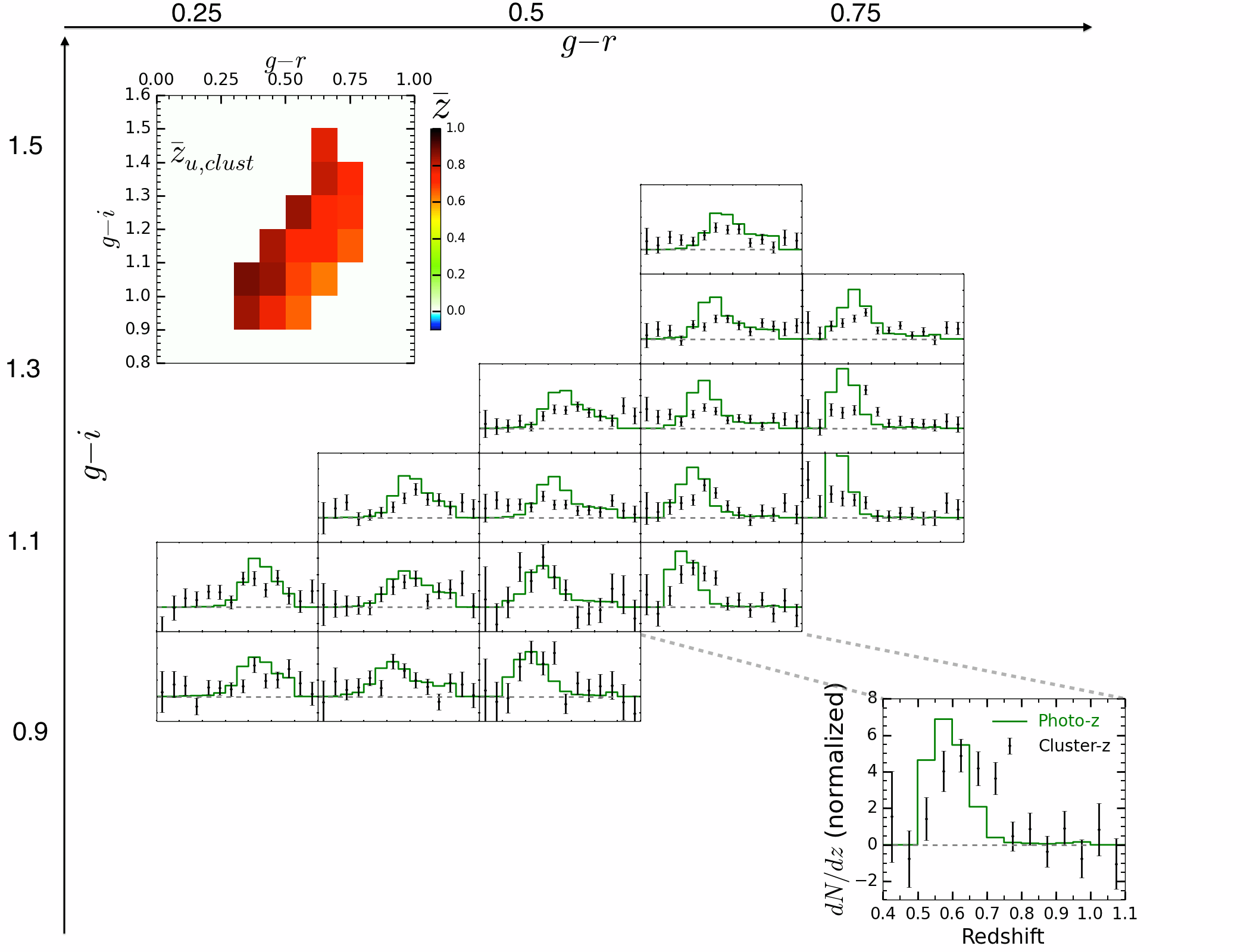}
\end{center}
\caption{Cluster-z (black points) and photo-z distributions in each cell of the all colour-space for $22.5< i_{\text{unk}} < 23$. The top left panel shows the evolution of the mean redshift with colours. A zoom for a given cell is shown in the bottom right panel.}
\label{fig_huge_map_225i23}
\end{figure*}

\noindent In this section we apply the same sampling approach in colour-space as previously but we use the fainter unknown sample defined in Section~\ref{sec:fainter_tomo} with $22.5 < i_{\text{unk}} < 23$.

\begin{figure*}
\begin{center}
\includegraphics[scale=0.42]{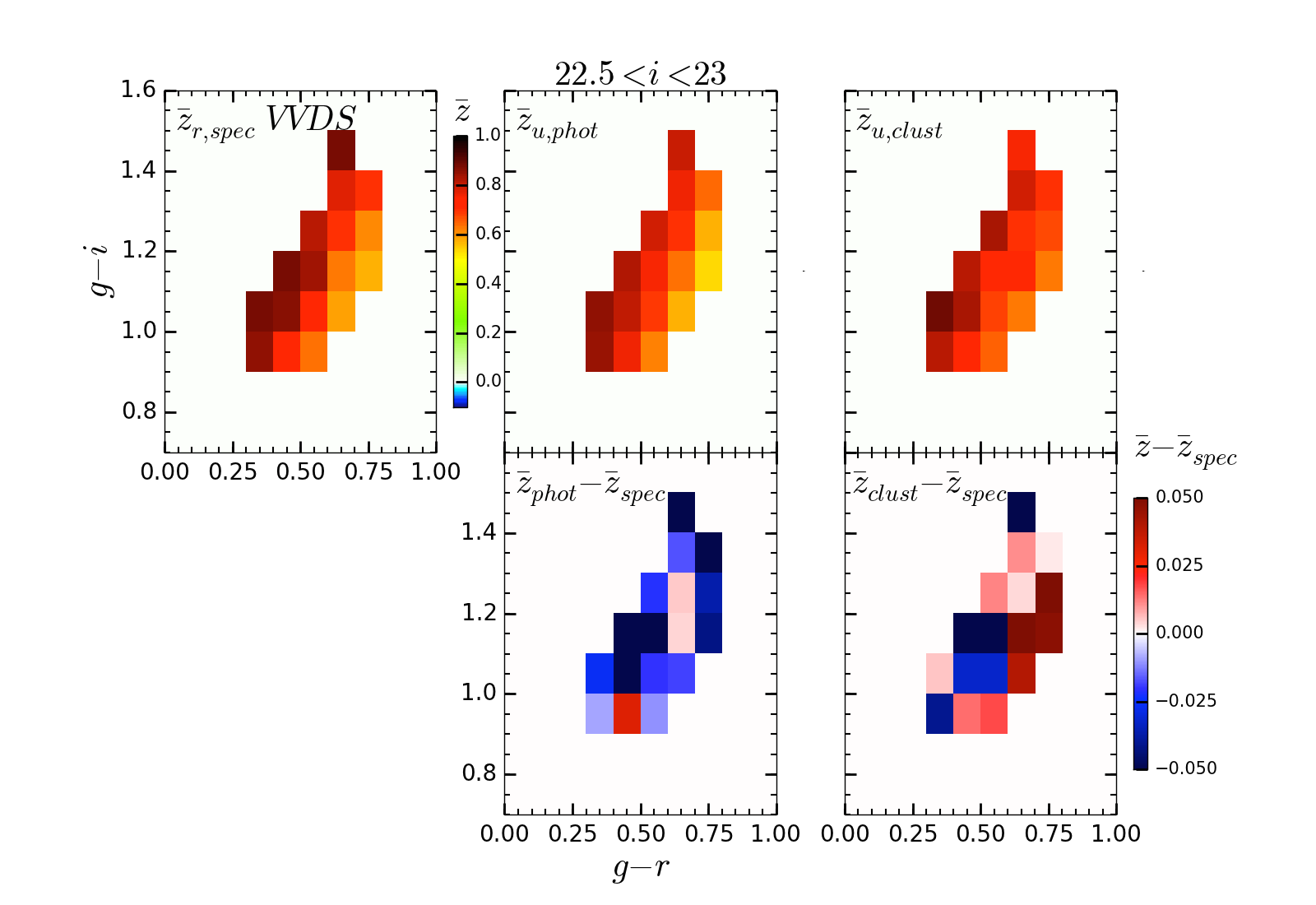}
\end{center}
\caption{\underline{TOP:} Mean redshift evolution through colour-space for the unknown sample and for spectroscopic data from VVDS. \newline
 \underline{BOT:} Difference between the estimated mean redshift \-- cluster or photo-z \-- and the true mean redshift from VVDS spectroscopic measurement. The residuals are all within the stochastic noise of the mean redshift estimate.}
\label{fig_fainter_mag_colors_maps}
\end{figure*}

\noindent This time the colour-redshift relation of both samples are not supposed to be the same. To check our results we then used VVDS data \citep{vvds_2005,vvds_2013} for which we corrected the magnitudes to be calibrated in the same way than the CFHTLS data. 

The VVDS data is then selected to have $22.5 < i < 23$ leading to a
complete sample of $\sim 1 \ 000$ sources. Since this sample is very
small and cover an area smaller than VIPERS, one can only expect to
have agreement on averaged statistical properties due to cosmic
variance.

Then we computed the corresponding clustering redshift distributions
visible in Figure~\ref{fig_huge_map_225i23}. Those distributions were
computed by considering $\deriv \beta_{\text{u}} / \deriv z = 1$.

~~\\

\noindent Figure~\ref{fig_fainter_mag_colors_maps} shows the resulting colour-redshift map in the top panels. The corresponding residuals for the faint colour sampling analysis are in the bottom panels. Due to the low number of spectroscopic sources, the residuals are all within the stochastic noise of the mean redshift estimate which can be estimated to be $\sim 0.1$. The summary statistics of these measurements are shown in Table~\ref{table_4}. 

In the top panel we found $\left\langle \Delta \bar{z}  \right\rangle = 0.03 $ and $\sigma= 0.05$ for photo-z \& $\left\langle \Delta \bar{z}  \right\rangle = 0.05 $ and $\sigma= 0.07$ for cluster-z when considering no evolution with redshift for $\beta_{\text{u}}$. While we found $\left\langle \Delta \bar{z}  \right\rangle = 0.04 $ and $\sigma= 0.06$ when considering $\deriv \beta_{\text{u}} / \deriv z = 1$, in the bottom panel. In both cases the cluster-z and photo-z measurements are in agreement.

Finally we combine all distributions from Figure~\ref{fig_huge_map_225i23} and reconstruct the global clustering redshift distribution of the fainter unknown population (see Figure~\ref{fig_fainter_mag_colors_sum_nz}). As we could expect we found results in good agreement between photo-z and cluster-z.

\begin{figure}
\begin{center}
\includegraphics[scale=0.34]{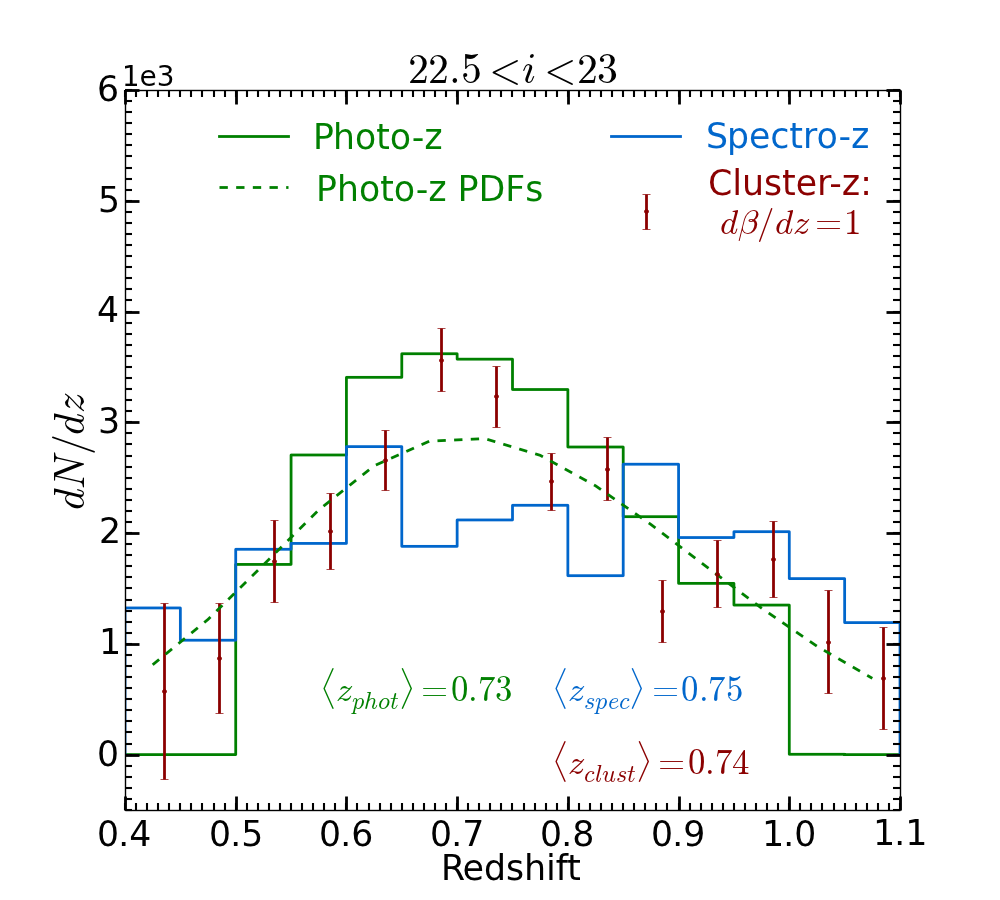}
\end{center}
\caption{Comparison between the global redshift distributions of the unknown sample measured by cluster-z with $\deriv \beta_{\text{u}} / \deriv z = 1 $ (red points), photo-z (green line), spectro-z from VVDS (blue line).}
\label{fig_fainter_mag_colors_sum_nz}
\end{figure}

\newpage

\noindent Figure~\ref{fig_fainter_mag_colors_sum_nz} shows that in the colour-sampling approach cluster-z and photo-z distributions are similar when the unknown sample is fainter than the population of reference.  As said previously it is difficult to make faint complete spectroscopic samples. This property of clustering-redshifts could be of great interest. Moreover we remind the reader that we use only 3 bands to subsample the unknown population whereas photo-z were obtained by combining optical and near-infrared data. We also remind that the spectroscopic distribution visible in blue is not the exact solution since it is a very small sample. Only averaged quatities should be compared. 

\newpage

\section{Summary}
\label{sec:concl}

We have explored  and quantified the ability of clustering-based redshift using VIPERS and CFHTLS. The method adopted in this paper follows the one presented in M13.

\begin{itemize}
\item  We demonstrated our ability to measure the clustering redshift distribution using a tomographic photo-z sampling. We found similar accuracy between photo-z and cluster-z. 

\item  We investigated the potential of this method to estimate redshift distributions of samples fainter than the magnitude limit of the reference population. In this case we also shown that the cluster-z accuracy is similar to photometric redshift. This suggest that the reference sample do not need to be representative of the unknown sample.  This property could be of great interest to estimate redshifts in the context of future large suveys.

\item  We have removed cluster-z from requirement of photo-z by selecting subsamples in colour-space. This allows the cluster-z measurement to be independant of the photometric redshift. That means that cluster-z does not suffer from possible systematics due to the photo-z procedure. Both methods could then be used to validate the other one. One could also try to combined them together.

\item We used the last two points to explore the ability of the clustering-based redshift estimation method to probe the redshift distribution of a sample in a magnitude range fainter than and non overlapping with
the reference population, independantly of photometric redshift. As said previously such property could be of great interest in the context of future large imaging surveys, like the Euclid space mission.
\end{itemize}

\noindent It is important to notice that in some case, e.g. for a
galaxy population with strong scale-dependent bias, the local approach
could not be sufficient. This would lead to a biased estimate of the
redshift distribution. Since the galaxy bias is a strictly increasing
function with redshift \citep{fry_1996,tegmark_1998} this would
induce an under/over estimation of the cluster-z distribution at
low/high redshift. Also cosmic variance can affect these results in
paricular in Section~\ref{sec:fainter_colors} when comparing cluster-z
to VVDS spectroscopic data.

In future works we will study in more detail within simulations the
accuracy reachable using this method in the context of Euclid. We will
also investigate the number of reference objects and number of filters
needed to reach the Euclid photo-z requirements, alone and/or when
combined with photo-z. Also the clustering properties of galaxies
beyond $z=1$ could affect the measurement. This will be explored in
future works. It is important to realise that the performance of this
approach will keep increasing, mostly because of the increase of the
spectroscopic data. Indeed, for a given unknown population, the
statistical noise will decrease with each new spectroscopic
redshift. And also because there is still plenty of information to be
added to break the colour-redshift degeneracy. For example, one can add
other kind of information such as size, brightness, ellipticity. These
points will also be explored in a future work.

\section*{Acknowledgments}

VS is particularly grateful to Brice M\'enard, without whom this paper would not have been possible.  His advices and many useful discussions have been essential for the development of this project. VS also thanks Mubdi Rahman for useful discussions. VS acknowledges funding from the French ministery for research, Universit\'e Pierre et Marie-Curie (UPMC) and the Centre National d'Etudes Spatiales (CNES) through the Convention CNES/CNRS N° 140988/00 on the scientific development of VIS \& NISP instruments and the management of the scientific consortium of the Euclid mission. VS acknowledges the Euclid Consortium and the Euclid Science Working Groups.

This work is also based on observations collected at the European Southern Observatory, Cerro Paranal, Chile, using the Very Large Telescope under programs 182.A-0886 and partly 070.A-9007. This paper is also based on observations obtained with MegaPrime/MegaCam, a joint project of CFHT and CEA/DAPNIA, at the Canada-France-Hawaii Telescope (CFHT), which is operated by the National Research Council (NRC) of Canada, the Institut National des Sciences de l’Univers of the Centre National de la Recherche Scientifique (CNRS) of France, and the University of Hawaii. This work is based in part on data products produced at Terapix available at the Canadian Astronomy Data Centre as part of the Canada-France-Hawaii Telescope Legacy Survey, a collaborative project of NRC and CNRS. The VIPERS web site is \url{http://www.vipers.inaf.it/} . This research makes use of the VIPERS-MLS database, operated at CeSAM/LAM, Marseille, France. VIPERS-MLS is supported by the ANR Spin(e) project (ANR-13-BS05-0005, \url{http://cosmicorigin.org}). This work is based in part on observations obtained with WIRCam, a joint project of CFHT, Taiwan, Korea, Canada and France. The TERAPIX team has performed the reduction of all the WIRCAM images and the preparation of the catalogues matched with the T0007 CFHTLS data release. This work is based in part on observations made with the Galaxy Evolution Explorer (GALEX). GALEX is a NASA Small Explorer, whose mission was developed in cooperation with the CNES of France and the Korean Ministry of Science and Technology. GALEX is operated for NASA by the California Institute of Technology under NASA contract NAS5-98034. This research uses data from the VIMOS VLT Deep Survey, obtained from the VVDS database operated by Cesam, Laboratoire d’Astrophysique de Marseille, France. 

We also acknowledge the crucial contribution of the ESO staff for the management of service observations. In particular, we are deeply grateful to M. Hilker for his constant help and support of this program. Italian participation to VIPERS has been funded by INAF through PRIN 2008 and 2010 programs. LG and BRG acknowledge support of the European Research Council through the Darklight ERC Advanced Research Grant (\# 291521). OLF acknowledges support of the European Research Council through the EARLY ERC Advanced Research Grant (\# 268107). AP, KM, and JK have been supported by the National Science Centre (grants UMO-2012/07/B/ST9/04425 and UMO-2013/09/D/ST9/04030), the Polish-Swiss Astro Project (co-financed by a grant from Switzerland, through the Swiss Contribution to the enlarged European Union). WJP and RT acknowledge financial support from the European Research Council under the European Community's Seventh Framework Programme (FP7/2007-2013)/ERC grant agreement n. 202686. WJP is also grateful for support from the UK Science and Technology Facilities Council through the grant ST/I001204/1. EB, FM and LM acknowledge the support from grants ASI-INAF I/023/12/0 and PRIN MIUR 2010-2011. LM also acknowledges financial support from PRIN INAF 2012. YM acknowledges support from CNRS/INSU (Institut National des Sciences de l'Univers) and the Programme National Galaxies et Cosmologie (PNCG). CM is grateful for support from specific project funding of the {\it Institut Universitaire de France} and the LABEX OCEVU. Research conducted within the scope of the HECOLS International Associated Laboratory, supported in part by the Polish NCN grant DEC-2013/08/M/ST9/00664.

\bibliographystyle{mn2e}
\bibliography{biblio}

\bsp

\newpage

\noindent $^{1}$Institut d’Astrophysique de Paris, UMR7095 CNRS, Universit\'e Pierre \& Marie Curie, 98 bis boulevard Arago, 75014 Paris, France\\
$^{2}$CEA/Irfu/SAp Saclay, Laboratoire AIM, F-91191 Gif-sur-Yvette, France\\
$^{3}$INAF - Osservatorio Astronomico di Brera, Via Brera 28, 20122 Milano, via E. Bianchi 46, 23807 Merate, Italy\\
$^{4}$INAF - Istituto di Astrofisica Spaziale e Fisica Cosmica Milano, via Bassini 15, 20133 Milano, Italy\\
$^{5}$Aix Marseille Universit\'e, CNRS, LAM (Laboratoire d'Astrophysique de Marseille) UMR 7326, 13388, Marseille, France\\
$^{6}$INAF - Osservatorio Astronomico di Torino, 10025 Pino Torinese, Italy\\
$^{7}$Aix Marseille Universit\'e, CNRS, CPT, UMR 7332, 13288 Marseille, France\\
$^{8}$INAF - Osservatorio Astronomico di Bologna, via Ranzani 1, I-40127, Bologna, Italy\\
$^{9}$Dipartimento di Matematica e Fisica, Universit\`{a} degli Studi Roma Tre, via della Vasca Navale 84, 00146 Roma, Italy\\
$^{10}$Institute of Cosmology and Gravitation, Dennis Sciama Building, University of Portsmouth, Burnaby Road, Portsmouth, PO1 3FX\\
$^{11}$Astronomical Observatory of the University of Geneva, ch. d'Ecogia  16, 1290 Versoix, Switzerland\\
$^{12}$INAF - Osservatorio Astronomico di Trieste, via G. B. Tiepolo 11, 34143 Trieste, Italy\\
$^{13}$Institute of Physics, Jan Kochanowski University, ul. Swietokrzyska 15, 25-406 Kielce, Poland\\
$^{14}$Dipartimento di Fisica e Astronomia - Alma Mater Studiorum Universit\`{a} di Bologna, viale Berti Pichat 6/2, I-40127 Bologna, Italy\\
$^{15}$INFN, Sezione di Bologna, viale Berti Pichat 6/2, I-40127 Bologna, Italy\\
$^{16}$Laboratoire Lagrange, UMR7293, Universit\'e de Nice Sophia Antipolis, CNRS, Observatoire de la C\^ote d’Azur, 06300 Nice, France\\
$^{17}$Astronomical Observatory of the Jagiellonian University, Orla 171, 30-001 Cracow, Poland\\
$^{18}$National Centre for Nuclear Research, ul. Hoza 69, 00-681 Warszawa, Poland\\
$^{19}$INAF - Istituto di Astrofisica Spaziale e Fisica Cosmica Bologna, via Gobetti 101, I-40129 Bologna, Italy\\
$^{20}$INAF - Istituto di Radioastronomia, via Gobetti 101, I-40129, Bologna, Italy\\
$^{21}$INFN, Sezione di Roma Tre, via della Vasca Navale 84, I-00146 Roma, Italy\\
$^{22}$INAF - Osservatorio Astronomico di Roma, via Frascati 33, I-00040 Monte Porzio Catone (RM), Italy

\label{lastpage}

\end{document}